\documentclass[12pt]{article}

\usepackage{a4wide, amsfonts, epsfig}
\newcommand{\be}{\begin{equation}} 
\newcommand{\ee}{\end{equation}}
\newcommand{\bea}{\begin{eqnarray}} 
\newcommand{\eea}{\end{eqnarray}}
\newcommand{\X}{\hbox{\rlap {$X$}\kern 2.2pt {$I$}}}
\newcommand{\noi}{\noindent} 
\newcommand{\np}{{}}
\newcommand{\ns}{\ \indent} 

\newcommand{\R}{{\bf R}}
\newcommand{\C}{{\bf C}}
\newcommand{\Z}{{\bf Z}}
\newcommand{\CP}{{\bf CP}}
\newcommand{\RP}{{\bf RP}}
\newcommand{\FC}{$F{\C}^3$}
\newcommand{\T}{{\cal T}}
\newcommand{\Y}{{\cal Y}}
\newcommand{\MD}{{\cal M}_{(2,[1])}}
\newcommand{\MM}[1]{{\cal M}_{(2,1)}^{#1}}
\newcommand{\M}{{\cal M}_{(2,1)}}
\newcommand{\MG}[1]{{\cal M}_{(#1)}}
\newcommand{\one}{$(\; ,1)$}
\newcommand{\onel}{$(1,\; )$}
\newcommand{\two}{$(2,\; )$}
\newcommand{\rone}{the hyperbolic region\/}
\newcommand{\rtwo}{the trigonometric region\/}
\newcommand{\sn}[2]{\,\mbox{sn}_#1\/#2}
\newcommand{\cn}[2]{\,\mbox{cn}_#1\/#2}
\newcommand{\dn}[2]{\,\mbox{dn}_#1\/#2}
\newcommand{\gn}[3]{\,\mbox{#3}_#1\/#2}
\newcommand{\inn}{\in\;}
\newcommand{\tr}{\mbox{trace}\,}
\newcommand{\hK}{hyperK\"ahler\ }
\newcommand{\mubf}{\mbox{\boldmath $\mu$}}

\newcommand{\news}{\setcounter{equation}{0}}

\begingroup\makeatletter\ifx\SetFigFont\undefined
\def\x#1#2#3#4#5#6#7\relax{\def\x{#1#2#3#4#5#6}}%
\expandafter\x\fmtname xxxxxx\relax \def\y{splain}%
\ifx\x\y   
\gdef\SetFigFont#1#2#3{%
  \ifnum #1<17\tiny\else \ifnum #1<20\small\else
  \ifnum #1<24\normalsize\else \ifnum #1<29\large\else
  \ifnum #1<34\Large\else \ifnum #1<41\LARGE\else
     \huge\fi\fi\fi\fi\fi\fi
  \csname #3\endcsname}%
\else
\gdef\SetFigFont#1#2#3{\begingroup
  \count@#1\relax \ifnum 25<\count@\count@25\fi
  \def\x{\endgroup\@setsize\SetFigFont{#2pt}}%
  \expandafter\x
    \csname \romannumeral\the\count@ pt\expandafter\endcsname
    \csname @\romannumeral\the\count@ pt\endcsname
  \csname #3\endcsname}%
\fi
\fi\endgroup

\begin{document}
\title{\vskip -70pt
  \begin{flushright}
    {\normalsize{
    DAMTP 98-5.\\ UDEM-GPP-TH 99-55.\\ Imperial/TP/98-99/022.\\hep-th/9902111.\\JHEP 04(1999)029.
    }} 
  \end{flushright}
  \vskip 15pt
  {\bf \Large \bf Two monopoles of one type and one of another}
  \vskip 10pt}
\author{ 
  Conor J. Houghton,\thanks{E-mail : C.J.Houghton@damtp.cam.ac.uk}
\\[5pt] 
{\normalsize {\sl Department of Applied Mathematics
      and Theoretical Physics,}}\\
  {\normalsize {\sl University of Cambridge, Silver St.,
      Cambridge, CB3 9EW, United Kingdom.}}\\[10pt]
Patrick W. Irwin,\thanks{E-mail : irwin@lps.umontreal.ca} 
\\[5pt]     
{\normalsize {\sl Groupe de Physique des Particules, D\'{e}partement de 
Physique,}} \\
{\normalsize {\sl Universit\'{e} de Montr\'{e}al,
C.P. 6128 succ. Centre-Ville,}} \\
{\normalsize {\sl Montr\'{e}al, Qu\'ebec, H3C 3J7, Canada.}}\\[10pt]
and
Arthur J. Mountain,\thanks{E-mail : A.Mountain@ic.ac.uk } \\[5pt]
  {\normalsize {\sl The Blackett Laboratory, Imperial College, Prince
Consort Road,}}\\ 
  {\normalsize {\sl London, SW7 2BZ, United Kingdom.}}\\ 
}

\date{February 1999}
 
\maketitle

\begin{abstract} 
  \noi The metric on the moduli space of charge (2,1) SU(3)
Bogo\-molny-Prasad-Sommer\-field monopoles is calculated and
investigated. The hyperKahler quotient construction is used to provide
an alternative derivation of the metric. Various properties of the
metric are derived using the hyperKahler quotient construction and the
correspondence between BPS monopoles and rational maps. Several
interesting limits of the metric are also considered.
\end{abstract}
\newpage
\section{Introduction}
\news
\ns This paper is about the metric on the moduli space of
$(2,1)$-monopoles. This metric is calculated and examined.
A $(2,1)$-monopole is a solution of the SU(3)
Bogomolny equation
\begin{equation}
D_i\Phi=B_i,
\end{equation}
where $D_i$ is the adjoint representation $\mathfrak{su}(3)$ covariant
derivative and $B_i$ is a nonAbelian magnetic field which is the Hodge
dual of the $\mathfrak{su}(3)$ field strength.  The Higgs field $\Phi$ is a
scalar field transforming under the adjoint representation of
$\mathfrak{su}(3)$. The Higgs field at infinity is required to lie in
the gauge orbit of
\begin{equation}
\Phi_{\infty}=i\left(\begin{array}{ccc}s_1&&
\\&s_2&\\&&s_3\end{array}\right), \label{P8} 
\end{equation} 
where $s_1+s_2+s_3=0$ and, by convention, $s_1<s_2<s_3$. This
condition on $\Phi$ gives a map from the large sphere at infinity into
the quotient space
\begin{equation} 
\mbox{orbit}_{\mbox{{\scriptsize
      SU}}(3)}\Phi_{\infty}=\mbox{SU}(3)/\mbox{U}(1)^2.
\end{equation} 
Since $\pi_2(\mbox{SU}(3)/\mbox{U}(1)^2) = {\bf Z}^2$, the moduli
space of monopoles is divided into sectors labelled by two topological
charges, $k_1$ and $k_2$, and a monopole with these charges is called a
$(k_1,k_2)$-monopole. These charges appear in the asymptotic expansion
of $\Phi$. For large $r=|{\bf x}|$, $\Phi$ lies in the gauge orbit of   
\begin{equation}
\Phi_\infty-\frac{i}{2r}\left(\begin{array}{ccc}k_1&&\\&k_2-k_1&\\&&-k_2\end{array}\right).\label{1overrF}
\end{equation}

$(k_1,0)$-monopoles are embeddings of
$k_1$-monopoles; SU(2) monopoles with topological charge $k_1$. The
embedding is essentially the trivial embedding of $\mathfrak{su}(2)$
into the upperleft $2\times2$ block of $\mathfrak{su}(3)$ \cite{B}.
The mass of a $(k_1,0)$-monopole is $4\pi Mk_1$, where

\begin{equation}
M=s_2-s_1.
\end{equation} 
In the same way, $(0,k_2)$-monopoles are $k_2$-monopoles with the
$\mathfrak{su}(2)$ embedded into the bottomright $2\times2$ block of
$\mathfrak{su}(3)$. The mass of a $(0,k_2)$-monopole is $4\pi mk_2$
where
\begin{equation}
m=s_3-s_2.
\end{equation}
In this way, there are two different types of SU$(3)$ monopoles: one
type corresponding to each U(1).  A $(k_1,0)$-monopole or a
$(0,k_2)$-monopole is made up of only one type of monopole and behaves
like the corresponding SU(2) monopole. A 
$(k_1,k_2)$-monopole has mass $4\pi(k_1M+k_2m)$. It is not
unreasonable to think of a $(2,1)$-monopole as being composed of two
monopoles of one type and one of another.

\np BPS monopoles interact in different ways depending on whether they
are of the same type or of different types. The metric on the moduli
space of two SU(3) monopoles of different type, the $(1,1)$-monopole,
is the Taub-NUT metric \cite{C,GL,LWY}. This metric can be derived
from physical arguments since it is the kinetic Lagrangian for the
electromagnetic and scalar interactions of point dyons. The metric for a
$(2,0)$-monopole, that is an SU(3) monopole with two monopoles of the
same type, is the Atiyah-Hitchin metric. This is the metric on the
moduli space of a 2-monopole.  When the monopoles are far apart, the
metric is exponentially close to a Taub-NUT metric which can be
interpreted as the kinetic Lagrangian for point dyons. When the
monopoles are not far apart, their interactions can not be understood
in this way. The metric on the moduli space of $(2,1)$-monopoles mixes
these interaction types.

\np In this paper, the metric on the moduli space of $(2,1)$-monopoles is
calculated from Nahm data. Nahm data are solutions to nonlinear matrix
equations in a single variable $s\in (s_1,s_3]$. The moduli space of
$(2,1)$ Nahm data is diffeomorphic to the moduli space of
$(2,1)$-monopoles. The calculation done in this paper is of the metric
on the moduli space of $(2,1)$ Nahm data. This is assumed to be
equivalent to calculating the metric on the monopole moduli space. We
will refer to this moduli space as $\M$.

\np The calculation is complicated and the resulting metric is not
transparent. Nonetheless it is possible to examine geodesic
submanifolds of the moduli space. By examining how the monopoles
behave in these submanifolds, it is possible to infer properties of the
monopole interactions. This is explained in Section \ref{submansect}.

\np During the calculation of the metric itself, it is useful to consider
some aspects of the corresponding physical picture. The metric on the
moduli space is the kinetic Lagrangian for three interacting monopoles
of two different types. It is possible to distinguish monopoles of
different types and it is also possible to think of the two different
monopole types separately. As explained below, each type corresponds to
separate, though interdependent, parts of the Nahm data. There is a
\two\ part of the Nahm data with $s\in(s_1,s_2]$ and a \one\ part
with $s\in[s_2,s_3]$. It is possible to think of these parts of the
Nahm data as corresponding to different part of the $(2,1)$-monopole,
a \one-monopole and a \two-monopole.

\np The \one\/ Nahm data are very simple; they consist of coordinates for
the \one-monopole.  The conditions on the Nahm data include a matching
condition at $s=s_2$. Because of this matching condition, the precise
form of the \two\/ Nahm data depends on the \one\/ Nahm data. This
means that
the precise shape of the \two-monopole depends on the position of
the \one-monopole. In fact, the further away the \one-monopole is, the
more the \two-monopole resembles a $(2,0)$-monopole.

\np What is remarkable is that up to group transformations, each
\two-monopole configuration corresponds to an ellipsoid of
\one-monopole positions. In other words, the matching condition and
gauge structure of the Nahm data is such that for a given class of
gauge equivalent \two\/ Nahm data there is an ellipsoid of \one\/ Nahm
data.  However, though each \two-monopole on this ellipsoid is
equivalent, the corresponding $(2,1)$-monopoles are all different,
since the \one-monopole is in a different position in each. This
difference disappears in the limit where the \one-monopole mass, $m$,
is zero. If $m=0$, the stabiliser of the asymptotic Higgs field
(\ref{P8}) is SU(2)$\times$U(1). This is the case of nonAbelian
residual symmetry.  This picture of nonAbelian residual symmetry as a
massless limit of the Abelian residual symmetry is due to Lee,
Weinberg and Yi \cite{LWY2}.

\np In the massless case, there is only one topological charge; there
is also a holomorphic charge which is preserved because of the complex
structure. The holomorphic charge counts the number of massless
monopoles.  We will distinguish between holomorphic and topological
charges by putting holomorphic charges in square brackets. In this
notation, the $m=0$ limit of a $(2,1)$-monopole is a
$(2,[1])$-monopole. In this space of $(2,[1])$-monopoles, 
repositioning the notional \one-monopole position is an isometry.  In
fact, this repositioning combined with the phase of the \one-monopole
constitutes an SU(2) isometric action on the $(2,[1])$-monopole moduli
space.

\np The metric on this moduli space is calculated by Dancer in
\cite{D}. It is a twelve-dimensional moduli space. These twelve
dimensions can
be interpreted as four dimensions parameterising the overall centre of
mass and position of the monopole, three dimensions for the SU(2)
isometry, three for the SO(3) orientation and, finally, two dimensions usually
called the separation and cloud parameters. These last two parameters
describe the separation of the two massive monopoles and the distance
from these to the ellipsoid of notional massless monopole
positions. This ellipsoid is often called the cloud.

\np The $(2,1)$ metric can be calculated in two parts. The first part
is due to the \two-monopole. This metric includes the effects of the
\one-monopole on the \two-monopole but does not include the
\one-monopole itself. The second part of the metric is due to the
\one-monopole. This is described in Section
\ref{datasect}. First, the Nahm data are introduced along with the
groups acting on them. Gauge-invariant coordinates on ${\cal
  M}_{(2,1)}$ are then defined from the Nahm data and a set of
one-forms defined from the exterior derivatives of these coordinates.
Tangent vectors dual to these one-forms are calculated and the metric
expressed in terms of the tangent vectors. The calculation of the
explicit metric is presented in Section \ref{metricsect}. In Section
\ref{HKQsect} another way of calculating the metric is discussed in
which the
physical description given above is very apparent.
In this Section, the \hK quotient construction is used to
construct the metric on $\M$ from the direct product of
the $(2,[1])$-monopole and 1-monopole metrics.

\np Section \ref{binet} contains a detailed discussion of the \hK
quotient construction in the context of the $(2,1)$ metric.
In Section \ref{propssect}, we derive certain properties of the moduli
space using rational maps. As mentioned above, Section \ref{propssect}
also contains an investigation of the geodesic submanifolds of $\M$
which correspond to sets of monopole configurations with extra
symmetry. Two asymptotic limits of the metric are calculated in Section
\ref{tam}. In Section \ref{l1msep}, the asymptotic expression is
calculated for large \one-monopole separation and in Section
\ref{ptdymet} the point dyon metric is discussed. Section
\ref{discusssect} contains concluding remarks. We explain that we were
motivated by two possible applications: Hanany-Witten theory and the
calculation of the approximate metric for the SU(2) three-monopole with
two monopoles close together and one far away. It is also noted that
there is an obvious variation on previous calculations, the
$([1],2,[1])$-monopole metric for SU(4) monopoles. This metric is
calculated in Section \ref{121sect}. 

\subsection{Notation and conventions}

\ns The calculations involve the use of different indices running over
different ranges. The following conventions are used.  Latin indices
$i$, $j$ and $k$ run from one to three.  Greek indices run from zero to
three or, in the case of $\lambda$, from one to four.  Latin indices
$a$, $b$ and $c$ are used for all other ranges, often one to eight or
one to seven.

\np We use two conventions for the generators of SU(2). The Pauli
matrices are written $\tau_i$ and satisfy $\tau_i \tau_j = \delta_{ij}
{\bf 1}_2 + i \epsilon_{ijk} \tau_k$, with ${\bf 1}_2$ the $2 \times 2$
unit matrix and $\tau_3={\rm diag}(1,-1)$. For simplicity of notation,
we also use the basis $e_i = -\frac{i}{2} \tau_i$ and the inner product
$\langle\; , \;\rangle = -2\tr (\; , \;)$. This gives $\langle
e_i,e_j\rangle=\delta_{ij}$ and $[e_i,e_j]=\epsilon_{ijk}e_k$.

\section{The moduli space of Nahm data}
\label{datasect}
\news

\subsection{The Nahm data}

\ns Nahm data are quite complicated to describe.  The Nahm data are matrix
functions of a single variable over the finite interval defined by the
eigenvalues of the asymptotic Higgs field. SU(3) Nahm data correspond to
the asymptotic Higgs field $\Phi_\infty$ in (\ref{P8}) and so the
interval is $(s_1,s_3]$. The interval is subdivided by the
intermediate eigenvalues. In our case, the interval is
subdivided by $s_2$ into $(s_1,s_2]$ and $[s_2,s_3]$.  Each
subinterval corresponds to a different monopole type and the dimension
of the associated matrix functions is equal to the charge of that type
of monopole. There are boundary conditions at the end points of the
interval and matching conditions between matrices at boundaries
between different subintervals. In each subinterval, the Nahm data satisfy
the Nahm equations.  From the Nahm data, the corresponding monopole
fields may be constructed \cite{Nah}.

\np The $(2,1)$ Nahm data are a quadruple $(\T_0,\T_1,\T_2,\T_3)$.  Each
$\T_{\mu}$ is a function of $s\in(s_1,s_3]$. For the lefthand
interval, $s\in (s_1,s_2]$, $\T_{\mu}=T_{\mu}$ a $2\times2$
skewHermitian matrix function satisfying the Nahm equation.  For the
righthand interval, $s\in [s_2,s_3]$, $\T_{\mu}=t_{\mu}$ is a
$1\times1$ skewHermitian matrix, in other words it is an
imaginary number. In the construction of monopole fields from Nahm
data, the \two\/ fields are derived from the $s\in (s_1,s_2]$ Nahm data
and the \one\/ fields from the $s\in [s_2,s_3]$ Nahm data. The reason
that the lefthand Nahm data are $2\times 2$ and the righthand Nahm data are
$1\times 1$ is that this is $(2,1)$ Nahm data. It can be represented by the
diagram
\begin{equation}
\begin{array}{c}
\begin{picture}(205,65)(55,615)
\thinlines
\put( 60,620){\line( 1, 0){200}}
\put(100,620){\line( 0, 1){ 40}}
\put(100,660){\line( 1, 0){ 60}}
\put(160,660){\line( 0,-1){ 20}}
\put(160,640){\line( 1, 0){ 70}}
\put(230,640){\line( 0,-1){ 20}}
\multiput(160,640)(0.00000,-8.00000){3}{\line( 0,-1){  4.000}}
\put( 95,623){\vector(0,1){  36}}  
\put( 95,623){\vector(0,-1){  0}} 
\put(230,620){\line( 0,-1){  5}}
\put(100,620){\line( 0,-1){  5}}
\put(160,620){\line( 0,-1){  5}}
\put( 84,638){\makebox(0,0)[lb]{\smash{\SetFigFont{12}{14.4}{rm}2}}}
\put( 97,605){\makebox(0,0)[lb]{\smash{\SetFigFont{12}{14.4}{rm}$s_1$}}}
\put(157,605){\makebox(0,0)[lb]{\smash{\SetFigFont{12}{14.4}{rm}$s_2$}}}
\put(227,605){\makebox(0,0)[lb]{\smash{\SetFigFont{12}{14.4}{rm}$s_3$}}}
\put(235,623){\vector(0,1){  15}}
\put(235,623){\vector(0,-1){  0}}
\put(240,626){\makebox(0,0)[lb]{\smash{\SetFigFont{12}{14.4}{rm}1}}}
\end{picture}\end{array}
\end{equation}

\np The Nahm equations are
\begin{equation}
  \frac{d \T_i}{ds} + [\T_0 , \T_i] = [\T_j , \T_k],
\end{equation}
where $(i\;j\;k)$ is a cyclic permutation of $(1\;2\;3)$. For the
$(2,1)$ Nahm data, this means that the lefthand Nahm data satisfy 
\begin{equation}
  \frac{d T_i}{ds} + [T_0 , T_i] = [T_j , T_k].\label{NE}
\end{equation}
On the righthand Nahm data, the Nahm equations mean that the $t_i$ are
constants. The boundary conditions require $T_i$ to have simple poles at
$s=s_1$ whose matrix residues form an irreducible representation of
$\mathfrak{su}(2)$. Finally, $\T_{\mu}$ is called continuous if 
\begin{equation}
t_{\mu} (s_2) = (T_{\mu}(s_2))_{2,2}. 
\label{contcond}
\end{equation} 
$\T_i$ are required to be continuous. $\T_0$ is not required to be
continuous.

\np In this paper, we use the Nahm data to calculate the metric on the
moduli space of monopoles. We do this by assuming that the moduli
space of monopoles and the moduli space of Nahm data are isometric; in
fact, this isometry has only been proven to exist for SU(2) \cite{Nak}.
The $L^2$ metric on the Nahm data is
\begin{eqnarray}\label{l2met}
  ds^2 &=&-\int_{s_1}^{s_3} \sum_\mu \tr (d\T_\mu d\T_\mu) ds\\
 &=&-\int_{s_1}^{s_2} \sum_\mu \tr (dT_\mu dT_\mu) ds -
    \int_{s_2}^{s_3} \sum_\mu dt_\mu dt_\mu ds,\nonumber
\end{eqnarray}
This is a hyperK\"ahler metric with hyperK\"ahler form
\begin{equation}
\omega= \int_{s_1}^{s_3} \tr{d\T\wedge\overline{d\T}}ds
=\int_{s_1}^{s_2}\tr{dT\wedge \overline{dT}}ds+\int_{s_2}^{s_3}
dt \wedge\overline{dt}
\end{equation}
where 
\begin{eqnarray}
d{\cal T}&=&d{\cal T}_0+Id{\cal T}_1+Jd{\cal T}_2+Kd{\cal T}_3,\\
\overline{d{\cal T}}&=&d{\cal T}_0-Id{\cal T}_1-Jd{\cal T}_2-Kd{\cal
  T}_3
\nonumber
\end{eqnarray}
and $(I,J,K)$ are quaternions.

\subsection{The Nahm construction}

\ns In order to construct monopole fields from Nahm data, the ADHMN
equation must be solved. This is the ordinary differential
equation
\begin{equation}
[{\bf 1}_{2k}\frac{d}{ds}+
({\bf 1}_k\otimes \sum_j x_j\sigma_j
+i\sum_j {\cal T}_j\otimes\sigma_j)]{\cal V}=0.
\end{equation}
For $s\in(s_1,s_2]$, $k=2$ and ${\cal V}=V(s;x_1,x_2,x_3)$, a complex
4-vector. $(x_1,x_2,x_3)$ is the point in space at which the field is
being constructed. For $s\in[s_2,s_3]$, $k=1$ and ${\cal V}=v(s;x_1,x_2,x_3)$,
a complex 2-vector. ${\cal V}$ is required to be continuous in the
sense that the three and four components of $V(s_2;x_1,x_2,x_3)$ are required
to be equal to $v(s_2;x_1,x_2,x_3)$. There is a three-dimensional
space of these ${\cal V}$ and an orthonormal basis is chosen with
respect to the inner product
\begin{equation}
({\cal V}_1,{\cal V}_2)=\int_{s_1}^{s_2}V_1^\dagger V_2ds+
\int_{s_2}^{s_3}v_1^\dagger v_2ds.  
\end{equation}
If ${\cal V}_1$, ${\cal V}_2$, ${\cal V}_3$ is such a basis the Higgs
field is 
\begin{equation}
(\Phi)_{ij}=(s{\cal V}_i,{\cal V}_j),
\end{equation}
with similar expressions for the gauge fields. 

In unpublished work done in collaboration with Paul M. Sutcliffe, this
construction has been performed for $(2,1)$-monopoles using the
numerical scheme of \cite{HS}. It is observed that, generally, there
appear to be three monopoles and that the \one-monopole energy peak
becomes lower and closer to the \two-monopole centre of mass as $m$ is
made smaller.

\subsection{Group actions on the Nahm data\label{gaNd}}
\label{groupactionsect}

\ns Another important aspect of the Nahm data is the group action. It
is also difficult to describe succinctly. Generally,
\begin{eqnarray}
  \T_0 &\mapsto& {\cal G} \T_0 {\cal G}^{-1}-\frac{d{\cal
    G}}{ds}{\cal G}^{-1}, \label{gact}\\
  \T_i &\mapsto& {\cal G} \T_i {\cal G}^{-1} \nonumber
\end{eqnarray}
defines an action on the Nahm data.
On the lefthand interval ${\cal G}$ is a U(2) matrix function $G$, on the
righthand interval it is a U(1) function $g$. Let us define
\begin{equation} 
G^{0,\star}=\{G\in C^\infty([s_1,s_2],\mbox{U}(2)):G(s_1)={\bf
  1}_2\},
\end{equation}
and 
\begin{equation} 
g^{\star,0}=\{g\in C^\infty([s_2,s_3],\mbox{U}(1)):g(s_3)=1\}.
\end{equation}
${\cal G}$ is called a gauge transformation if ${\cal G}\in{\cal
  G}^{0,0,0}$ where
\begin{equation} 
{\cal G}^{0,0,0}=\left\{{\cal G}=(G,g): G\in G^{0,\star},\; g\in
  g^{\star,0},\;
  G(s_2)=\left(\begin{array}{cc}1&0\\0&g(s_2)\end{array}\right)
\right\} .
\end{equation}
It should be noted that the gauge transformation satisfies a strong
condition at $s=s_2$. As well as requiring continuity
$(G(s_2))_{2,2}=g(s_2)$ the condition also fixes the other entries of $G(s_2)$.

\np This action is called a gauge action, because monopoles constructed
from gauge equivalent Nahm data are gauge equivalent.  The space of
Nahm data is only diffeomorphic to the monopole moduli space, once the
gauge action has been factored out. This factored space is called the
moduli space of Nahm data.

\np The reason for writing the gauge group ${\cal G}^{0,0,0}$ with
three superscripted zeros is that there are three boundary and
junction conditions beyond the continuity condition
$(G(s_2))_{2,2}=g(s_2)$. The first zero refers to the $G(s_1)$ being
the identity, the second zero to the strong condition at $s=s_2$:
\begin{equation}
G(s_2)=\left(\begin{array}{cc}1&0\\0&g(s_2)\end{array}\right),
\end{equation}
and the third zero to $g(s_3)$ being the identity.
For similar groups in which the second or third condition is relaxed,
the corresponding zero is replaced by a star. These groups are
important, because group transformations which do not satisfy these two
conditions nonetheless map Nahm data to Nahm data and are used
frequently in the calculation of the metric. They are used to derive
all the Nahm data from a specific ansatz solution and the group
parameters are used as coordinates on the moduli space.
Thus, the groups we will need are
\begin{equation} 
{\cal G}^{0,\star,0}=\left\{{\cal G}=(G,g): G\in G^{0,\star},\; g\in
  g^{\star,0},\;
  (G(s_2))_{2,2}=g(s_2)
\right\}
\end{equation}
and
\begin{equation} 
{\cal G}^{0,\star,\star}=\left\{{\cal G}=(G,g): G\in G^{0,\star},\; g\in
  g^{\star,\star},\; (G(s_2))_{2,2}=g(s_2) \right\},
\end{equation}
where
\begin{equation} 
g^{\star,\star}=\{g\in C^\infty([s_2,s_3],\mbox{U}(1))\}.
\end{equation}
We will also use
\begin{equation} 
G^{0,0}=\{G\in C^\infty([s_1,s_2],\mbox{U}(2)):G(s_1)={\bf
  1}_2,\;G(s_2)={\bf 1}_2\}.
\end{equation}
The group ${\cal G}^{\star,0,0}$, in which the first condition is
relaxed, is used later in this Section when defining the rotational
SO(3) action.

\np In the $(2,[1])$ case considered by Dancer, the asymptotic Higgs
field has two equal eigenvalues, this means
\begin{equation} 
\Phi_{\infty}=i\left(\begin{array}{ccc}s_1& &
    \\&s_2&\\&&s_2\end{array}\right),
\end{equation}
which has little group U(2). The $(2,[1])$ case has minimal symmetry
breaking; SU(3) is broken to U(2).  Since $m=0$, the Nahm data may be
represented as 
\begin{equation}
\begin{array}{c}
\begin{picture}(205,65)(55,605)
\thinlines
\put( 60,620){\line( 1, 0){130}}
\put(100,620){\line( 0, 1){ 40}}
\put(100,660){\line( 1, 0){ 60}}
\put(160,660){\line( 0,-1){ 20}}
\put(160,640){\line( 1, 0){  2}}
\put(162,640){\line( 0,-1){ 20}}
\multiput(160,640)(0.00000,-8.00000){3}{\line( 0,-1){  4.000}}
\put( 95,623){\vector(0,1){  36}}  
\put( 95,623){\vector(0,-1){  0}} 
\put(162,620){\line( 0,-1){  5}}
\put(100,620){\line( 0,-1){  5}}
\put(160,620){\line( 0,-1){  5}}
\put( 84,638){\makebox(0,0)[lb]{\smash{\SetFigFont{12}{14.4}{rm}2}}}
\put( 97,605){\makebox(0,0)[lb]{\smash{\SetFigFont{12}{14.4}{rm}$s_1$}}}
\put(157,605){\makebox(0,0)[lb]{\smash{\SetFigFont{12}{14.4}{rm}$s_2$}}}
\put(167,623){\vector(0,1){  15}}
\put(167,623){\vector(0,-1){  0}}
\put(172,626){\makebox(0,0)[lb]{\smash{\SetFigFont{12}{14.4}{rm}1}}}
\end{picture}\end{array}
\end{equation}
The gauge action on this Nahm data is given by $G^{0,0}$ and the
unbroken U(2) is $G^{0,\star}/G^{0,0}$. In the $(2,[1])$ moduli space
this U(2) is an isometry. There is a similar U(2) action when $m$ is not
zero. However, it is not an isometry. This U(2) action is ${\cal
  G}^{0,\star,\star}/{\cal G}^{0,0,0}$. The U(2) isometry found in the
$(2,[1])$ case has split into two parts. There is an S$^3$ part at
$s=s_2$ which may be written as the coset ${\cal G}^{0,\star,0}/{\cal
  G}^{0,0,0}$ and a U(1) which may be written as the coset ${\cal
  G}^{0,0,\star}/{\cal G}^{0,0,0}$.

\np It is useful to discuss the physical meaning of the group
transformations. The \one-monopole has a well defined
position given by 
\begin{equation}
{\bf r}=-i(t_1,t_2,t_3).\label{lgpos}
\end{equation}
This is fixed under the maximal torus of the U(2)$={\cal
  G}^{0,\star,\star}/{\cal G}^{0,0,0}$ action consisting of group
elements which are diagonal at $s=s_2$. The action of this maximal
torus is isometric. The orbit of ${\bf r}$ under the U(2) action is an
ellipsoid; this will be clear when we write down solutions of the Nahm
equations in Section \ref{solvingNe}. This ellipsoid corresponds to
the coset U(2)$/$U(1)$^2$. Thus the group action consists of an
isometric maximal torus and a nonisometric coset which moves the
\one-monopole around on an ellipsoid. 

\np There is also an SO(3) action which rotates the whole
monopole in space. The Nahm data are acted on by rotating
$(\T_1,\T_2,\T_3)$ as a three-vector.
 Precisely, if $A\in {\cal G}^{\star,0,0}$
and $A(s_1)$ maps to the SO(3) matrix $(A_{ij})$, this action is
defined by
\begin{eqnarray}\label{SO3action}
  T_0&\rightarrow&AT_0A^{-1}-\frac{dA}{ds}A^{-1},\\
  T_i&\rightarrow&A(\,\sum_j A_{ij}T_j\,)A^{-1},\nonumber\\
  t_0&\rightarrow&t_0-\frac{1}{a}\frac{da}{ds},\nonumber\\
  t_i&\rightarrow&\sum_j A_{ij}t_j,\nonumber
\end{eqnarray}
where $(A(s_2))_{2,2}=a(s_2)$. Thus, the SO(3) action rotates the
$(\T_1,\T_2,\T_3)$ without disturbing the matrix residues at $s=s_1$.

Finally, there is a translational $\R^3$ action on the Nahm data. It is
given on the Nahm data by 
\begin{eqnarray}
T_i&\mapsto&T_i+i\lambda_i{\bf 1}_2, \\
t_i&\mapsto&t_i+i\lambda_i.\nonumber\label{transact}
\end{eqnarray}
This action is an isometry and corresponds to a translation in space
of the whole monopole.

\subsection{The centre of mass}
\label{CoMs}

\ns The overall motion of a monopole is free and the (2,1) moduli space
decomposes as 
\begin{equation}
\M=\R^3\times\frac{\R\times\M^0}{\Z},
\end{equation}
where the $\R^3$ and $\R$ correspond to the overall position and overall
phase. $\M^0$ is called the centred moduli space and contains the
nontrivial structure of $\M$. Calculating the metric on $\M^0$ is the
main business of this paper and this calculation is simplified by the
observation that the traces of the $T_\mu$ can be set to zero. This is
explained later in this Section.

\np It is easy to see from the Nahm equations (\ref{NE}) that the
traces of the $T_i$ are constant. This is something they have in
common with the righthand data; $t_i$ are also constant. A gauge can
be chosen so that both $\tr{T_0}$ and $t_0$ are constants. This does
not define a specific gauge or specific constants: neither constant is
gauge invariant. The invariant combination is $M\tr{T_0}+mt_0$ and
this combination plays a significant role in the calculation done in
this Section.

The metric can now be rewritten in terms of traceless data and the
traces themselves: the Nahm data $T_\mu$ are replaced by
$T_\mu+iR_\mu{\bf 1}_2$ where $T_\mu$ is now traceless. Defining {\bf
  r} as above, (\ref{lgpos}), and $r_0=-it_0$, the metric (\ref{l2met})
becomes 
\begin{equation} 
  ds^2 = 2M \sum_\mu dR_\mu dR_\mu + m \sum_\mu dr_\mu dr_\mu
   -\int_{s_1}^{s_2} \sum_\mu \tr d T_\mu d T_\mu ds.
\label{metric2}
\end{equation}

\begin{figure}[t]
\begin{center}
\leavevmode
\epsfig{file=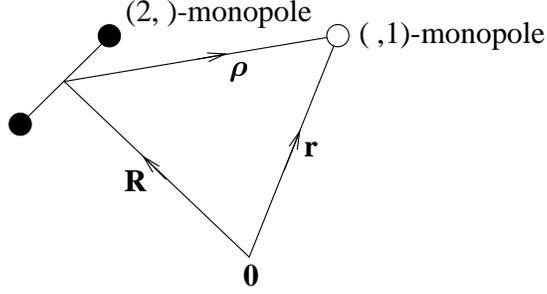,bbllx=195pt,bblly=385pt,bburx=400pt,bbury=500pt,
width=205pt}
\caption{An illustration of the definition of the various separation vectors.}
\end{center}
\label{crfig}
\end{figure}

\np Furthermore, the action of the translation action and the fact that the
\one-monopole position is ${\bf r}$, suggests that the centre of mass
of the \two-monopole is positioned at ${\bf R}$. This is
illustrated in Figure 1, along with the relative position
vector 
\begin{equation}
  \rho_i = r_i-R_i.
  \label{newri}
\end{equation}
This implies that the centre of mass is given by 
\begin{equation}
  \frac{2M R_i + m r_i}{2M+m} = R_i + \frac{m}{2M+m} \rho_i
\end{equation}
and, in fact,
\begin{equation} \label{splitmetric}
  ds^2 =ds^2_{\mbox{centre}}+ \frac{2Mm}{2M+m} \sum_\mu
  d\rho_\mu d\rho_\mu - \int_{s_1}^{s_2}\sum_\mu \tr
  dT_\mu dT_\mu ds,
\end{equation}
where $\rho_0 = R_0-r_0$ and
\begin{equation}
ds^2_{\mbox{centre}}=(2M+m)d\left(\frac{2M R_\mu + m
      r_\mu}{2M+m}\right)
d\left(\frac{2M R_\mu + m r_\mu}{2M+m}\right).
\end{equation}
Thus, the metric separates into two terms, one of which is flat. In
Section \ref{metricsect} the metric on the term which is not flat is
calculated. The $\rho_i$ satisfy the same continuity conditions with
respect to the traceless Nahm data as $r_i$ do with respect to the general
Nahm data. This means that the metric we want is the metric on the space
of traceless Nahm data. 

Thus, in Section \ref{metricsect} the metric is calculated on the
space of Nahm data with traceless $T_i$ and $T_0$ and with the mass $m$
replaced by the reduced mass $2Mm/(2M+m)$ as in (\ref{splitmetric}). For
simplicity, rather than changing to the reduced mass, we will continue
to use $m=s_3-s_2$ and to denote the space by $\M^0$. In fact, it is
demonstrated above that in the metric on $\M^0$ the mass $m$ must be
replaced by the reduced mass. 

The fixing of the centre of mass is discussed from a different
perspective in Section \ref{binet}, where it is fixed using a
hyperK\"ahler quotient.

\subsection{Solving the Nahm equations\label{solvingNe}}

\ns The Nahm equations on the lefthand interval are easy to solve. The
ansatz 
\begin{eqnarray}
&&T_0(s)=0, \label{basicdata} \\
&&T_i(s)= -\frac{i}{2}f_i(s) \tau_i = f_i(s) e_i,\nonumber
\end{eqnarray}
reduces them to the well-known Euler-Poinsot equations 
\begin{equation} 
\frac{d}{ds}f_i(s)=f_j(s)f_k(s),
\end{equation}
where $(i\;j\;k)$ is an cyclic permutation of $(1\;2\;3)$.
Assuming $[f_1(s)]^2\le [f_2(s)]^2\le [f_3(s)]^2$ the solutions are the Euler top functions 
\begin{eqnarray}\label{topfunctions}
  f_1(s) & = & -\frac{D \cn{k}{D(s-s_1)}}{\sn{k}{D(s-s_1)}},\\
  f_2(s) & = & -\frac{D \dn{k}{D(s-s_1)}}{\sn{k}{D(s-s_1)}},\nonumber\\
  f_3(s) & = & -\frac{D}{\sn{k}{D(s-s_1)}}.\nonumber
\end{eqnarray}
These solutions have a pole of the correct form at $s=s_1$. $k$ is the
elliptic parameter and $0\le k\le 1$. In order for the Nahm data to be
nonsingular inside the interval $(s_1,s_2]$, we must have $D<2K(k)/M$,
where $K(k)$ is the usual complete elliptic integral of the first
kind. Thus, the ansatz yields a two-parameter space of solutions. The
Nahm data $t_i$ on the righthand interval are determined by the
continuity condition (\ref{contcond}) at $s=s_2$.

\np Of course, not all Nahm data are produced by this ansatz. However,
all the required Nahm data can be produced by acting on this
two-parameter ansatz space with the group actions described in Section
\ref{groupactionsect} and below. In other words, $D$ and $k$ label
group orbits and the complete orbit of the ansatz space is the whole
manifold. The orbits are not identical, however, and the ansatz space
is not a manifold.

There are two group actions on the Nahm data. First, there is the SO(3)
rotation action given in (\ref{SO3action}). On uncentred Nahm data, we
also have an action of ${\cal G}^{0,\star, \star}/{\cal G}^{0,0,0} =
\left(\mbox{U}(2) \times \mbox{U}(1) \right)/\mbox{U}(1)$. One effect of
centring the Nahm data by the procedure of Section \ref{CoMs} is that the
group action must have unit determinant. Thus, the correct group action
on the centred Nahm data is SO(3)$\times \mbox{S}\left({\cal
G}^{0,\star, \star}/{\cal G}^{0,0,0} \right) = \mbox{SO}(3)\times
\mbox{SU}(2)$. After a general transformation of this type, the Nahm data can
be put in the form
\begin{eqnarray}
  \label{generalpoint}
  ({T_0},{T_i}) &=& \left( -G\frac{dA}{ds} A^{-1}G^{-1}-\frac{dG}{ds}
    G^{-1},GA(\sum_j A_{ij}f_j(s)e_j)A^{-1}G^{-1} \right) ,\\
  ({t}_0,{t}_i) &=& \left( -\frac{da}{ds}a^{-1}-\frac{dg}{ds}
    g^{-1},(T_i(s_2))_{2,2} \right). \nonumber
\end{eqnarray}
This is an eight-dimensional space of Nahm data. This
eight-dimensional space is assumed to be isometric to the relative
moduli space of $(2,1)$-monopoles. We use $\M^0$ to denote both of
these moduli spaces.

\subsection{Coordinates and one-forms}
\label{1formsect}

\ns Eight coordinates on $\M^0$ are now required. In order to define
such coordinates, an explicit representation of the group action at
$s=s_2$ in terms of Euler angles is needed. This is given by
\begin{equation}
  G(s_2) = \left( {\begin{array}{cc} 
    \cos{\frac{\theta}{2}} e^{-\frac{i}{2}(\chi+\phi)} & 
    -\sin{\frac{\theta}{2}} e^{\frac{i}{2}(\chi-\phi)} \\
    \sin{\frac{\theta}{2}} e^{-\frac{i}{2}(\chi-\phi)} &
    \cos{\frac{\theta}{2}} e^{\frac{i}{2}(\chi+\phi)} \end{array}} \right).
\end{equation}
Since the group action on the Nahm data is an adjoint action, it
descends from SU(2) to SO(3) action under the usual homomorphism 
\begin{equation}
  G(s_2) \mapsto E_{ij}=\frac{1}{2}\tr \left( \tau_i G(s_2) \tau_j
  G(s_2)^\dagger \right). 
\end{equation}
In terms of the Euler angles, the matrix $E$ is 
\begin{equation}
\label{eulerangs}
\left( \begin{array}{lll}
    \cos \theta \cos \chi \cos \phi - \sin \phi \sin \chi & 
    - \cos \theta \cos \phi \sin \chi - \cos \chi \sin \phi &
    \cos \phi \sin \theta \\
    \cos \phi \sin \chi + \cos \theta \cos \chi \sin \phi &
    - \cos \theta \sin \phi \sin \chi + \cos \chi \cos \phi &
    \sin \theta \sin \phi \\
    - \cos \chi \sin \theta & 
    \sin \theta \sin \chi &
    \cos \theta
  \end{array} \right).
\end{equation}
The Nahm data are acted on by both this group action and the rotation action
defined in (\ref{SO3action}). After acting with general elements of
these groups, the traceless Nahm data are
\begin{eqnarray}
{T}_1(s)&=& \sum_{i,j} A_{1i} f_i(s) E_{ji} e_j,\\
{T}_2(s)&=& \sum_{i,j} A_{2i} f_i(s) E_{ji} e_j,\nonumber\\
{T}_3(s)&=& \sum_{i,j} A_{3i} f_i(s) E_{ji} e_j.\nonumber
\end{eqnarray}

Five of the coordinates
introduced by Dancer on the (2,[1]) moduli space can be
immediately adopted as coordinates on $\M^0$.
\begin{eqnarray}
  \alpha_1 &=& \langle {T}_1,{T}_1 \rangle-\langle
    {T}_2,{T}_2 \rangle,\\
  \alpha_2 &=& \langle {T}_1,{T}_1 \rangle-\langle
    {T}_3,{T}_3 \rangle,\nonumber \\
  \alpha_3 &=& \langle {T}_1,{T}_2 \rangle,\nonumber\\
  \alpha_4 &=& \langle {T}_1,{T}_3 \rangle,\nonumber \\
  \alpha_5 &=& \langle {T}_2,{T}_3 \rangle.\nonumber 
\end{eqnarray}
These coordinates are invariant under the gauge and group actions and
independent of $s$. $\alpha_3$, $\alpha_4$ and $\alpha_5$ are
coordinates for the rotational action. In the $(2,[1])$ moduli space
combinations of $\alpha_1$ and $\alpha_2$ are the separation and cloud
parameters.  They play a similar role here, except that the cloud
parameter in the $m=0$ case becomes a genuine separation parameter
when $m$ is not zero. It then corresponds to the separation of the
\one-monopole from the centre of the \two-monopole. The last three
coordinates used in \cite{D} are
\begin{eqnarray}
  \cos{\alpha_6} & = & \frac{\langle {T}_3,e_3 \rangle}{\| 
    {T}_3 \|}, \\
  \cos{\alpha_7} & = & \frac{\langle [{T}_3,{T}_2],
    [e_3,{T}_3] \rangle}{\|[{T}_3,{T}_2]\|
    \|[e_3,{T}_3]\|}, \nonumber \\
  \sin{\alpha'_8} & = & \frac{\langle {T}_3,e_2
    \rangle}{\|[{T}_3,e_3]\|},\label{ttc}
\end{eqnarray}
where, in each case, the righthand side is evaluated at $s=s_2$.

Two of these coordinates, $\alpha_6$ and $\alpha_7$, are
gauge-invariant, that is, invariant under a general element of ${\cal
  G}^{0,0,0}$. If $A={\bf 1}_3$ they correspond to the $\theta$ and
$\chi$ Euler angles which determine the position of the \one-monopole
on the ellipsoid. Thus they parameterise an $S^2$ surface acted on by
SU(2). 

The third coordinate, $\alpha_8'$, is not gauge invariant.
The diagonal U(1) subgroup of the SU(2) action corresponds to the
relative phase of the two monopoles. The overall phase of the
\two-monopole is given by $\alpha'_8$. The expression for the phase of a
$(\ldots,\;, 1,\; ,\ldots)$-monopole in the moduli space of $(1,1,\ldots,
1)$-monopoles is $i \int {t}_0 ds$ \cite{Mu}. Consequently, we try the
combination

\begin{equation}
  \label{relphase}
  \alpha_8={\alpha}_8'+i \int_{s_2}^{s_3} {t}_0 ds.  
\end{equation}
This is a well-defined coordinate: direct calculation shows that it is
gauge-invariant. It has a clear physical interpretation; the two terms
are the phases of the \two-monopole and the \one-monopole in that
order. This shows that the relative phase of the \one-monopole and the
\two-monopole is changed by a group action but not by a gauge
action. Furthermore, the normalisation of $\alpha_8$ is correct because
it is identified modulo $2\pi$ under the group action ${\cal
G}^{0,0,\star}/{\cal G}^{0,0,0}$. 

It should be noted that the coordinates above reduce to the
coordinates used by Dancer \cite{D} in the $m \rightarrow 0$
limit. This is a smooth limit. 

\subsection{One-forms and tangent vectors}
\label{tangentsect}

\ns Now that we have a full set of coordinates for $\M^0$ we can compute
the metric. To do this, a basis of one-forms is derived by taking the
exterior derivatives of the coordinates: $\alpha_a$, $a=1\ldots8$.  A
basis of tangent vectors orthogonal to these one-forms is then found
and their inner product calculated. This gives a local expression for
the metric.

\np Thus, it is now important to define a basis of physically
motivated and computationally convenient one-forms and express the
metric in terms of these.  One-forms corresponding to $\alpha_1$ to
$\alpha_5$ are defined as $\rho_a = d \alpha_a$ for $a=1\ldots 5$. To
deal with the SU(2) group action we start by considering the set of
left-invariant one-forms.  These are
\begin{eqnarray}
  \rho_6 &=& \cos{\alpha_8'} d\alpha_6+ \sin{\alpha_8'} \sin{\alpha_6}
    d\alpha_7, \\
  \rho_7 &=& -\sin{\alpha_8'} d\alpha_6+ \cos{\alpha_8'} \sin{\alpha_6} d
    \alpha_7, \nonumber \\
  \rho_8' &=& d\alpha_8' + \cos{\alpha_6} d\alpha_7. \nonumber
\end{eqnarray}
To perform the explicit calculation of the metric, matters are
simplified greatly if we use the U(1) isometry to work at infinitesimal
$\alpha_8'$. It may be assumed that for all explicit calculations, we
are working in this limit in which, neglecting infinitesimal terms,
these expressions become

\begin{eqnarray} \label{oneforms}
  \rho_6&=& d\alpha_6, \nonumber \\
  \rho_7&=& \sin{\alpha_6} d \alpha_7,\nonumber\\
  \rho_8'&=& d\alpha_8' + \cos{\alpha_6} d\alpha_7.\nonumber
\end{eqnarray}
The reason $\rho_8'$ appears with a prime is that we will have cause
to modify it when we come to calculate the metric. We reserve the
symbols $\rho_a$ for $a=1\ldots 8$ for the one-forms used in the
calculation. 

\np $\rho_8'$ is well-defined.  Although
$\alpha_8=\alpha_8'+i\int_{s_2}^{s_3}t_0ds$ rather than either
$\alpha_8'$ or $i\int_{s_2}^{s_3}t_0ds$ alone is the well-defined
coordinate, $d\alpha_8' + \cos\alpha_6 d\alpha_7$ and
$d(i\int_{s_2}^{s_3}t_0ds)$ are both well-defined one-forms.  They each
give gauge-invariant results when contracted with tangent vectors. The
tangent vectors are discussed in the remainder of this Section and
the gauge invariance of the contraction can be seen by direct
calculation in the case of $d(i\int_{s_2}^{s_3}t_0ds)$. In the case of
$d\alpha_8'+\cos\alpha_6 d\alpha_7$ gauge invariance follows from the
fact it is the one-form used in the calculation of the $(2,[1])$
metric and this metric is isometric under $G^{0,\star}$.

\np The next problem is that of finding the tangent vectors to a point in
$\M$. In the Nahm description, these tangent vectors are quadruples
$(\Y_0,\Y_1,\Y_2,\Y_3)$ where each $\Y_\mu$ is a $2\times2$
skewHermitian matrix function, $Y_\mu$, on the lefthand interval and
an imaginary function, $y_\mu$, on the righthand interval. In other
words, $Y_\mu(s) \in {\mathfrak u}(2)$ and $y_\mu(s) \in {\mathfrak
  u}(1)$. In order for these vectors to
be tangent to $\M$, $\Y_i$ must satisfy the linearisation of the
conditions satisfied by $\T_i$. This means that they satisfy the linearised
Nahm equations. On the lefthand interval, the linearised Nahm equations
are
\begin{eqnarray}
  \label{tangentconds}
  \frac{dY_1}{ds}+[Y_0,{T}_1]+[{T}_0,Y_1]&=&[{T}_2,Y_3] 
    +[Y_2,{T}_3], \\
  \frac{dY_2}{ds}+[Y_0,{T}_2]+[{T}_0,Y_2]&=&[{T}_3,Y_1] 
+[Y_3,{T}_1],\nonumber \\
  \frac{dY_3}{ds}+[Y_0,{T}_3]+[{T}_0,Y_3]&=&[{T}_1,Y_2] 
    +[Y_1,{T}_2]. \nonumber
\end{eqnarray}
On the righthand interval the linearised Nahm equations require $y_i$ to
be constant. To preserve the tracelessness of the lefthand Nahm data
$\tr Y_i$ must be set to zero. This is consistent since
(\ref{tangentconds}) implies that $\tr Y_i$ are constant. Furthermore,
the $\Y_i$ must be continuous. This means that we have $y_i =
\left(Y_i(s_2)\right)_{2,2}$. Since the points of $\M$ correspond to
gauge-equivalent sets of Nahm data, the tangent vectors must also be
orthogonal to the gauge transformations.  The inner product on the
tangent space is given by 
\begin{equation} 
  ({\cal X},{\cal Z})=-\int_{s_1}^{s_2}\sum_\mu \tr X_\mu Z_\mu ds
    -\int_{s_2}^{s_3}\sum_\mu x_\mu z_\mu ds,
\label{innerpr}
\end{equation}
where ${\cal X}$ and ${\cal Z}$ are two tangent vectors composed of
the quadruples $({\cal X}_0,{\cal X}_1,{\cal X}_2,{\cal X}_3)$ and 
$({\cal Z}_0,{\cal Z}_1,{\cal Z}_2,{\cal Z}_3)$. 
The metric is derived from this inner product and it
is with respect to this inner product that the tangent vectors and the
gauge transformations are required to be orthogonal. An infinitesimal
gauge transformation has the form
\begin{equation}
 \left(\frac{d
 \Psi}{ds}+[{T}_0,\Psi],[{T}_1,\Psi],[{T}_2,\Psi],
  [{T}_3,\Psi] \right),
\end{equation}
on the lefthand interval. $\Psi$ is a $\mathfrak{u}(2)$ function which
is zero at $s=s_1$ and proportional to diag$(0,1)$ at $s=s_2$. On the
righthand interval, the infinitesimal transformation has the form
\begin{equation} 
\left(\frac{d\psi}{ds},0,0,0\right),
\end{equation}
where $\psi$ and $\Psi$ satisfy the continuity condition $\psi(s_2) =
\left(\Psi(s_2) \right)_{2,2}$. The
orthogonality conditions derived from this transformation are
\begin{eqnarray}
  \frac{dY_0}{ds}+\sum_\mu [{T}_\mu,Y_\mu]&=&0,\\
  \frac{dy_0}{ds}&=&0,\nonumber
\end{eqnarray}
and $\Y_0$ must be continuous:
\begin{equation}
  y_0=\left(Y_0(s_2)\right)_{2,2}.
\end{equation}
It is noteworthy that gauge orthogonality imposes similar conditions
on $\Y_0$ as those already imposed on $\Y_i$ by the linearised Nahm
equations. This is related to the hyperK\"ahlerity of $\M$. 

It is useful that the conditions above are similar to those found by
Dancer \cite{D}.  The conditions on the $Y_\mu$ are precisely the
same. The additional conditions do not impose additional constraints;
they dictate the values of $y_\mu$ corresponding to a given $Y_\mu$.
This means that the tangent vectors at the point where $(T_0,T_1,T_2,T_3)$
is $(0,f_1(s) e_1,f_2(s) e_2,f_3(s) e_3)$ are known; they follow from
those presented in \cite{D}. For the sake of completeness, these are
reproduced here. $Y_{\mu}(s)$ are elements of the Lie algebra of SU(2)
and so can be represented by $Y_{\mu}(s)=\sum_{i=1}^3 Y_{\mu
  i}(s)e_i$.  Representing the index $i$ as a vector index the
solutions of the linearised Nahm equation are 
\begin{eqnarray}\label{tvectors}
Y_0(s)=\left(\begin{array}{ccc}\dot{f_1}(s)I_4\\\dot{f_2}(s)I_3+m_3/f_2(s)\\
    -\dot{f_3}(s)I_2-n_2/f_3(s)\end{array}\right),\qquad
Y_1(s)=\left(\begin{array}{ccc}\dot{f_1}(s)I_1\\\dot{f_2}(s)I_2+m_2/f_2(s)\\
    \dot{f_3}(s)I_3+n_3/f_3(s)\end{array}\right),\\
Y_2(s)=\left(\begin{array}{ccc}-\dot{f_1}(s)I_2\\\dot{f_2}(s)I_1+m_1/f_2(s)\\
    -\dot{f_3}(s)I_4-n_4/f_3(s)\end{array}\right),\qquad
Y_3(s)=\left(\begin{array}{ccc}-\dot{f_1}(s)I_3\\\dot{f_2}(s)I_4+m_4/f_2(s)\\
    \dot{f_3}(s)I_1+n_1/f_3(s)\end{array}\right),\nonumber 
\end{eqnarray} 
where
\begin{equation}
I_\lambda(s)=m_\lambda g_1(s)+n_\lambda g_2(s),
\end{equation}
and $m_\lambda$, $n_\lambda$ are real numbers which
parametrise the eight dimensional tangent space. $g_1$ and $g_2$ are
the incomplete elliptic integrals
\begin{eqnarray}\label{iei}
    g_1(s)&=&\int_{s_1}^{s} \frac{1}{f_2(s)^2} ds, \\
    g_2(s)&=&\int_{s_1}^{s} \frac{1}{f_3(s)^2} ds.\nonumber
\end{eqnarray}
The tangent vectors transform in a
simple fashion under the SO(3) and SU(2) group actions. Tangent vectors
at the general point in $\M^0$ given by (\ref{generalpoint}) are
 \begin{equation}
  \{{\Y}_0, {\Y}_i \} = \{ G \Y_0 G^{-1} , G A_{ij} \Y_j
    G^{-1} \}.
\end{equation}
There are eight independent tangent vectors at every point of $\M^0$.
Given a set of eight coordinates $\alpha_a$ on $\M^0$, a
basis of eight tangent vectors given in components as
${\Y}_\mu^a$ can be defined by
\begin{equation} \label{orthonorm}
  d \alpha_a ({\Y}^{b}) = \lim_{\epsilon \rightarrow 0} \left(
  \frac{\alpha_a({\T}_\mu+\epsilon {\Y}^{b}_\mu) -
  \alpha_a({\T}_\mu)}{\epsilon}\right) = \delta^b_a. 
\end{equation}
\noi This method is then used to construct tangent vectors orthonormal
to the one-forms.

\subsubsection{The dual bases of one-forms and tangent vectors}

\ns In order to calculate the metric, we need a basis of one-forms and
a basis of tangent vectors such that the two bases are dual in the
sense of (\ref{orthonorm}). We wish to replace the one-form $\rho_8'$
with a one-form which is derived from the well-defined angle $\alpha_8$
rather then $\alpha_8'$. However, if we proceed in the straightforward
manner using the basis of one-forms derived from the coordinates
$\rho_a$  for $a=1\ldots 7$ and 
\begin{equation} \label{trialrho8}
\rho_8'+d\left(i \int_{s_2}^{s_3} {t}_0 ds
\right)=d\alpha_8+\cos{\alpha_6}d\alpha_7,
\end{equation}
we create unnecessary computational difficulties. The
expressions for the tangent vectors which result are both very
complicated and very different from those used by Dancer to calculate
the $(2,[1])$ metric.  Instead, we apply the orthonormalisation
procedure as follows. We begin by defining the eight tangent vectors
used by Dancer \cite{D} in the calculation of the $(2,[1])$
metric. These are $\Y'^a$ for $a=1\ldots8$ which are dual to $\rho_a$
for $a=1\ldots7$ and to $\rho'_8$. Thus 
\begin{equation}
  \rho_a(\Y'^b) = \delta_a^b,
\end{equation}
for $a=1\ldots 7$ and $b=1\ldots8$ and
\begin{equation}
  \rho'_8(\Y'^b) = \delta_8^b,
\end{equation}
for $b=1\ldots8$. As these are the $(2,[1])$ tangent vectors, the
orthonormalisation depends on $Y'^a$ and not on $y'^a$. 

We must now introduce $\rho_8$, which is the modified form of
$\rho_8'$. If we proceed in the straightforward manner illustrated in
(\ref{trialrho8}) and we use
\begin{equation}
d \left( i\int_{s_2}^{s_3} t_0 ds \right)(\Y'^a)=imy'^a_0,
\end{equation}
we see that the one-form (\ref{trialrho8}) is not dual to the convenient
set of tangent vectors. Instead we define 
\begin{equation}\label{rho8}
  \rho_8 = d\alpha_8 +
    \cos{\alpha_6} d\alpha_7-im\sum_{a=1}^7 y^a_0\rho_a. 
\end{equation}
With this new eighth one-form $\rho_8(\Y'^a)$ is zero for
$a=1\ldots7$. Clearly, the conditions $\rho_a(\Y'^b)= \delta_a^b$ for
$a,b=1 \ldots 7$ are unaffected. Also, $\rho_8$ is a good one-form as it
involves only the exterior derivatives of well-defined coordinates. It
only remains to look at the action of $\rho_8$ on $\Y'^8$. This is
\begin{equation}\label{Wscale}
\rho_8(\Y'^8)=\Omega,
\end{equation}
where 
\begin{equation}
\Omega=1+imy'^8_0.
\end{equation}
Thus, the basis of dual tangent vectors is given by $\Y^a$ where
\begin{equation}
{\Y}^a=\Y'^a,
\end{equation}
for $a=1\ldots 7$ and 
\begin{equation}
{\Y}^8=\frac{1}{\Omega}\Y'^8.
\end{equation}
What this means is that $\rho_8$ is proportional to $\rho'_8=
d\alpha'_8 + \cos\alpha_6 d\alpha_7$, in fact, 
\begin{equation} \label{Wscale2}
\rho_8 = \Omega \rho'_8 = \Omega(d\alpha'_8 + \cos\alpha_6 d\alpha_7).
\end{equation} 
An explicit expression for $\Omega$ is given later.  Note that, apart
from this change to the eighth one-form and eighth tangent vector, the
one-forms and tangent vectors are precisely those used in the
construction of the $(2,[1])$ metric \cite{D, I}.  The eight tangent
vectors can be written out in terms of the coordinates $m_\lambda$ and
$n_\lambda$ above but it would be tedious to give these expressions
here. 

In summary, the one-form $\rho_8'$ is well-defined but is not identified
modulo $2\pi$ under the group action and cannot therefore be written in
terms of Euler angles. We have constructed a one-form $\rho_8$ in terms
of well-defined coordinates which turns out to be simply a rescaling of
$\rho_8'$.

\subsection{The metric}

\ns The tangent vectors are uniquely determined by the
orthonormalisation procedure in Section \ref{tangentsect} and we can
now construct the metric on $\M^0$ in terms of the inner products of
these tangent vectors. The contribution to the metric from the
lefthand Nahm data is given by the formula, found in \cite{D},
\begin{equation}
  ds^2_{(2,\;)} = -\int_{s_1}^{s_2} \sum_{a,b=1}^8 \sum_{\mu}
    \tr{{Y}_\mu^{a}  {Y}_\mu^{b}} ds \, \rho_a \rho_b.
\label{metricpart1}
\end{equation}
The contribution to the metric from the righthand Nahm data is given
by a similar formula
\begin{equation}
  ds^2_{(\;,1)} = -\int_{s_2}^{s_3} \sum_{a,b=1}^8 \sum_{\mu}
    {y}_\mu^a {y}_\mu^b ds \, \rho_a \rho_b = - \int_{s_2}^{s_3} \sum_{\mu}
    d {t}_\mu d {t}_\mu . 
  \label{metricpart2}
\end{equation}
The explicit expression for the metric is given in the next Section.

\section{The explicit metric}
\label{metricsect}
\news 

\ns To calculate the explicit metric, it is convenient to follow
the version of the $m=0$ metric calculation given in \cite{I}. In
$\MD^0$ the SU(2) group action and the SO(3) rotational action are both
isometric. This allows the metric to be calculated at one point on the
SU(2)$\times$SO(3) orbit; there is a two-parameter space of such
orbits. For the $(2,1)$ metric, the isometric
actions on $\M^0$ are the U(1) action of ${\cal G}^{0,0,\star}/{\cal
G}^{0,0,0}$ and the SO(3) rotational action. The righthand Nahm data,
corresponding to a \one-monopole at ${\bf r}$, break the SU(2) group
isometry to a U(1) isometry. This means that the relative metric must be
calculated on the whole four-dimensional space of SO(3)$\times$U(1)
orbits. This space is parameterised by two relative separations and two
Euler angles giving the relative orientation of the monopoles. The
isometry is used to calculate the metric for an infinitesimal
SO(3)$\times$U(1) action which will give the metric on the whole of
$\M^0$. 

\np Working at a point infinitesimally close to the identity on the
SO(3)$\times$U(1) orbit means that the coordinates $\alpha_6$,
$\alpha_7$ and $\alpha_8'$ are equal to the Euler angles $\theta$, $\chi$
and $\phi$ respectively, up to infinitesimal terms in the SO(3)
action. However, when the derivative of one of these coordinates is
taken, these extra terms will, in general, contribute non-trivial
couplings of the SO(3) one-forms to $d\theta$, $d\chi$ and
$d\phi$. This is shown explicitly in (\ref{explicit1forms}). Of course
the gauge-dependent quantities $\alpha_8'$ and $\phi$ do not appear in
the metric, although, as discussed in Section \ref{tangentsect}, their
exterior derivatives are well-defined one-forms on the moduli space and
do appear in the metric.

\subsection{The contribution from the lefthand Nahm data}
\ns The notation used in \cite{D, I} is adopted here. 
Much of the calculation involves evaluating the elliptic functions
$f_i(s)$ at $s=s_2$ and so, in the rest of the paper, if there is no
explicit argument given, it is assumed that the value at $s=s_2$ is
used, that is $f_i=f_i(s_2)$. The same notation is used for the
elliptic integrals (\ref{iei}): $g_1$ is used to mean $g_1(s_2)$ and
$g_2$ to mean $g_2(s_2)$. Of course, these integrals are still
incomplete elliptic integrals, even though they are integrals over the
whole of $(s_1,s_2]$.  It is also convenient to introduce the
combinations
\begin{eqnarray}
    X&=&f_1 f_2 f_3,\\
    p_1&=&g_1+\frac{1}{X},\nonumber\\ 
    p_2&=&g_2+\frac{1}{X} ,\nonumber\\
    p_3&=&g_1+g_2+\frac{1}{X}.\nonumber
\end{eqnarray}
In this notation, the scale $\Omega$ introduced earlier (\ref{Wscale}) is
\begin{equation}
\Omega=1+\frac{m}{2}\left[\frac{X(g_1+g_2)}{f_1^2p_3}\sin^2\theta
    \cos^2\chi +\frac{Xp_1}{g_1f_2^2}\sin^2\theta \sin^2\chi
    +\frac{Xp_2}{g_2f_3^2}\cos^2\theta\right].
\end{equation}
It is the nor\-mal\-i\-sa\-tion of the one-form $\rho_8$.

In terms of the
elliptic function parameters $D$ and $k$, the Euler angles
(\ref{eulerangs}) on the ellipsoid and the SO(3) one-forms
$\{\sigma_i\}$ defined by  
\begin{equation}
  A^\dagger d A |_{s=s_1} = \frac{i}{2} \sum_{i} \tau_i
  \sigma_i,
\end{equation}
explicit expressions for our set of one-forms are 
\begin{eqnarray} \label{explicit1forms}
  \rho_1 &=& d \alpha_1 = d(-k'^2D^2), \\
  \rho_2 &=& d \alpha_2 = d(-D^2), \nonumber\\
  \rho_3 &=& d \alpha_3 = -k'^2D^2 \sigma_3, \nonumber\\
  \rho_4 &=& d \alpha_4 = -D^2 \sigma_2, \nonumber\\
  \rho_5 &=& d \alpha_5 = k^2D^2 \sigma_1, \nonumber \\
  \rho_6 &=& d \alpha_6 = d\theta +\frac{f_2}{f_3} \sin{\chi}
    \sigma_1 +\frac{f_1}{f_3} \cos{\chi} \sigma_2, \nonumber\\
  \rho_7 &=& \sin{\alpha_6} d\alpha_7 = \sin{\theta} d\chi
    +\frac{f_2}{f_3} \cos{\theta} \cos{\chi} \sigma_1 -\frac{f_1}{f_3}
    \cos{\theta} \sin{\chi} \sigma_2 +\frac{f_1}{f_2} \sin{\theta}
    \sigma_3, \nonumber\\
  \frac{1}{\Omega}\rho_8 &=& d\alpha'_8 + \cos{\alpha_6}
    d\alpha_7 = d\phi +\cos{\theta}
    d\chi -\frac{f_2}{f_3} \sin{\theta}\cos{\chi} \sigma_1
    +\frac{f_1}{f_3} \sin{\theta} \sin{\chi} \sigma_2 +\frac{f_1}{f_2}
    \cos{\theta} \sigma_3 . \nonumber
\end{eqnarray}
$k^\prime$ is the dual modulus: $k^{\prime 2}= 1- k^2$. These
expressions for the one-forms hold at a point on the SO(3)$\times$U(1)
orbit.

It is clear that any linearly-independent set of one-forms constructed
from these could be suitable one-forms and could be used to describe
the metric. The set of one-forms that lead to the simplest expression
for the metric and most emphasise the connection to the $(2,[1])$
metric \cite{I} are
\begin{eqnarray}
  \kappa_1 &=&-\sin\chi \rho_6 -\cos\chi \cos\theta \rho_7
    + \sin\theta \cos\chi \frac{1}{\Omega} \rho_8,\\
  \kappa_2 &=&-\cos\chi \rho_6 + \sin\chi \cos\theta \rho_7
    - \sin\theta\sin\chi \frac{1}{\Omega} \rho_8, \nonumber\\
  \kappa_3 &=& -\sin\theta \rho_7 -  \cos\theta \frac{1}{\Omega} \rho_8
    .\nonumber 
  \label{newoneforms}
\end{eqnarray}
In terms of these, the contribution to the metric from the traceless
part of the lefthand Nahm data can be written
\begin{eqnarray}
  ds^2_{(2,\;)}&=&\frac{1}{8}\left[X(g_1d\alpha_1+g_2d\alpha_2)^2
+ g_1d\alpha_1^2 + g_2d\alpha_2^2\right] \\
&+&\frac{1}{2}\left(a_1 \sigma_1^2 +a_2 \sigma_2^2 +a_3 \sigma_3^2\right) \nonumber \\
&+&\frac{1}{2}\left[\left(b_1\sigma_1+c_1\kappa_1\right)^2+\left(b_2\sigma_2+c_2\kappa_2\right)^2
+\left(b_3\sigma_3+c_3\kappa_3\right)^2\right],\nonumber
\end{eqnarray}
where
\begin{equation}
  a_1=\frac{g_1g_2 k^4 D^4}{g_1+g_2}, \qquad 
  a_2=\frac{g_2p_3 D^4}{p_1} , \qquad
  a_3=\frac{p_3g_1 D^4k^{\prime 4}}{p_2},
\end{equation}
\begin{equation}
  b_1=-k^2D^2\sqrt{\frac{g_2^2}{X(g_1+g_2)p_3}},\qquad
  b_2=\frac{g_2D^2}{\sqrt{Xg_1p_1}},\qquad
  b_3=\frac{g_1D^2k^{\prime 2}}{\sqrt{Xg_2p_2}},
\end{equation}
and
\begin{equation}
  c_1=\frac{1}{f_1}\sqrt{\frac{X(g_1+g_2)}{p_3}},\qquad
  c_2=\frac{\sqrt{Xg_1p_1}}{f_2 g_1}, \qquad
  c_3=\frac{\sqrt{Xg_2p_2}}{f_3g_2}.
\end{equation}

\subsection{The contribution from the righthand Nahm data}

\ns Next, let us turn to the contributions to the metric from the
righthand Nahm data. Inevitably, these have rather long expressions
as it is this part of the metric which breaks the isometry from SU(2)
to U(1). In the limit of large separation, where the SU(2) isometry is
restored, these expressions simplify. The contribution from
$t_1$, $t_2$ and $t_3$ is
\begin{equation}
  ds^2 = -\int_{s_2}^{s_3} \sum_{i} \sum_{a,b=1}^7 {y}_i^{a}
    {y}_i^{b} \rho_a \rho_b ds = m d {\bf r}\cdot d {\bf r}. 
\end{equation}
These Nahm data are functions of only the spatial position coordinates
$\alpha_a$ where $a=1 \ldots 7$ and, consequently, the metric contribution
will only involve $\rho_a$ where $a=1\ldots 7$. The vector ${\bf r}$ is
known explicitly, it is 
\begin{equation}
  {\bf r} = \frac{1}{2} A\left( \begin{array}{c} -f_1 \sin{\theta} \cos{\chi} \\ f_2 \sin{\theta} \sin{\chi} \\ f_3 \cos{\theta} \end{array} \right), 
  \label{ti}
\end{equation}
where $A$ is the SO$(3)$ matrix $A_{ij}$ defined in (\ref{SO3action}).
Differentiating this expression gives the one-form $d{\bf r}$ as
\begin{eqnarray}
2d r_1 &=&\frac{1}{2} f_2 f_3\sin{\theta} \cos{\chi} (g_1 d
  {\alpha}_1 +g_2 d {\alpha}_2)-\frac{1}{f_3}\cos{\theta}
  D^2\sigma_2 \\
& &+\frac{1}{f_2} \sin{\theta} \sin{\chi} D^2 k^{\prime 2} \sigma_3 +f_1
  \cos{\theta} \kappa_2 -f_1 \sin{\chi} \sin{\theta} \kappa_3, \nonumber \\
2d r_2 &=&-\frac{1}{2}f_1 f_3\sin{\theta} \sin{\chi} (p_1 d
  {\alpha}_1 +g_2 d {\alpha}_2)
  +\frac{1}{f_3}k^2D^2\cos{\theta} \sigma_1 -f_2\cos{\theta} \kappa_1
  \nonumber \\
& &-f_2\cos{\chi} \sin{\theta} \kappa_3, \nonumber \\
2d r_3 &=& -\frac{1}{2}f_1 f_2{\cos} \theta (g_1 d
  {\alpha}_1 +p_2 d {\alpha}_2)+f_3 \sin{\theta}\sin{\chi} \kappa_1 +f_3 \sin{\theta}\cos{\chi} \kappa_2.\nonumber
\end{eqnarray}
The final term in the metric is associated
with changes of the phase of the \one-monopole. This is
\begin{equation}
  \frac{1}{m} \left[d(i\int_{s_2}^{s_3}t_0 ds)\right]^2 
    = -m \sum_{a,b=1}^8 {y}_0^{a} {y}_0^{b} \rho_a \rho_b. 
\end{equation}
where
\begin{eqnarray}
  2i\sum_{a=1}^8 {y}_0^{a} \rho_a 
 &=&  -\frac{(g_1+g_2)X}{p_3f_1^2}\sin\theta \cos\chi\kappa_1+\frac{p_1X}{g_1f_2^2}
  \sin\theta \sin\chi \kappa_2 +\frac{p_2X}{g_2f_3^2} \cos\theta
  \kappa_3\\
 &+& 
\frac{g_2}{p_3f_1}k^2D^2\sin\theta \cos\chi
  \sigma_1 +\frac{g_2}{g_1f_2} D^2 \sin\theta \sin\chi \sigma_2 +
  \frac{g_1}{g_2f_3}D^2k^{\prime 2}\cos\theta \sigma_3.
\nonumber
\end{eqnarray}

\subsection{The complete metric}

\ns For completeness, the whole metric is now displayed. It is
\begin{eqnarray}  \label{fullmetric}
ds^2_{(2,1)}&=&\frac{1}{8}\left[X(g_1d\alpha_1+g_2d\alpha_2)^2
+ g_1d\alpha_1^2 + g_2d\alpha_2^2\right] \\
&+&\frac{1}{2}\left(a_1 \sigma_1^2 +a_2 \sigma_2^2 +a_3 \sigma_3^2\right) \nonumber \\
&+&\frac{1}{2}\left[\left(b_1\sigma_1+c_1\kappa_1\right)^2+\left(b_2\sigma_2+c_2\kappa_2\right)^2
+\left(b_3\sigma_3+c_3\kappa_3\right)^2\right] \nonumber \\
&+&\frac{m}{4}\left[\frac{1}{2} f_2 f_3\sin{\theta} \cos{\chi} (g_1 d
  {\alpha}_1 +g_2 d {\alpha}_2)-\frac{1}{f_3}\cos{\theta}
  D^2\sigma_2 \right. \nonumber \\
& & \left. +\frac{1}{f_2} \sin{\theta} \sin{\chi} D^2 k^{\prime 2} \sigma_3
  +f_1 \cos{\theta} \kappa_2 -f_1 \sin{\chi} \sin{\theta} \kappa_3
  \right]^2 \nonumber \\
&+&\frac{m}{4}\left[-\frac{1}{2}f_1 f_3\sin{\theta} \sin{\chi} (p_1 d
  {\alpha}_1 +g_2 d {\alpha}_2) \right. \nonumber\\
& &\left. +\frac{1}{f_3}k^2D^2\cos{\theta} \sigma_1 -f_2\cos{\theta}
  \kappa_1-f_2\cos{\chi} \sin{\theta} \kappa_3 \right]^2 \nonumber \\
&+&\frac{m}{4}\left[-\frac{1}{2}f_1 f_2\cos \theta (g_1 d
  {\alpha}_1 +p_2 d {\alpha}_2)+f_3 \sin\theta\sin\chi
  \kappa_1 +f_3 \sin\theta\cos\chi \kappa_2 \right]^2 \nonumber \\
&+&\frac{m}{4}\left[\frac{g_2}{p_3f_1}k^2D^2\sin\theta \cos\chi
  \sigma_1 +\frac{g_2}{g_1f_2} D^2 \sin\theta \sin\chi \sigma_2 +
  \frac{g_1}{g_2f_3}D^2k^{\prime 2}\cos\theta\sigma_3 \right. \nonumber\\
& &\left. -\frac{(g_1+g_2)X}{p_3f_1^2}\sin\theta \cos\chi \kappa_1
  +\frac{p_1X}{g_1f_2^2}
  \sin\theta \sin\chi \kappa_2 +\frac{p_2X}{g_2f_3^2} \cos\theta
  \kappa_3 \right]^2.\nonumber
\end{eqnarray}

This is the metric on the space of traceless Nahm data. To derive the
metric on $\M$, the mass $m$ must be replaced by the reduced mass and
the overall centre of mass term
\begin{equation}
(2M+m)\sum_\mu dR_\mu dR_\mu
\end{equation}
must be added.

In the $m \rightarrow 0$ limit $\Omega=1$ and $\{\kappa_i\}$ become a
basis of body-fixed one-forms for the isometric SU(2) action on $\MD$.
Bearing this in mind, it is easy to see that $ds^2_{(2,1)}$ reduces to
$ds^2_{(2,[1])}$ when $m=0$. Some of the properties of the metric are
examined in Section \ref{propssect}. The asymptotic metric which
results when the \one-monopole is a great distance from the
\two-monopole is calculated in Section \ref{l1msep}.

\section{The \hK quotient construction and $\M$}
\label{binet}
\news

\ns The quotient of a \hK manifold by a group action is not
hyperK\"ahler. In the \hK quotient construction \cite{HKLR}, the
moment map is used to restrict the manifold to a level set whose
quotient is hyperK\"ahler. In short, if there is a free isometric
group action of a compact group $G$ on a \hK manifold ${\cal M}$ which
preserves the complex structures, there is a triplet of moment maps
\begin{equation}
  \mu_i:{\cal M}\mapsto \mathfrak{g}^\star.
\end{equation} 
The contraction of the moment map $\mu_i$ with an element $\xi$ of
$\mathfrak{g}$ is a function on ${\cal M}$. The exterior derivative of
this function is identical to the contraction of the K\"ahler form
$\omega_i$ with the Killing vector field corresponding to $\xi$. The
manifold
\begin{equation}
{\cal N}=\mu^{-1}({\bf c})/G
\end{equation}
is a \hK manifold provided $c_1$ $c_2$ and $c_3$ are central elements
of $\mathfrak{g}^\star$.

The Nahm equations appear naturally in the \hK quotient construction
because they are moment maps for the gauge action. The moduli
space of Nahm data is the \hK quotient of the space of general matrix
functions with the correct boundary conditions by the gauge group
\cite{Hi}. This was exploited by Murray in his
calculation of the $(1,1,\ldots,1)$ metric \cite{Mu}. 

In this Section, we are concerned with U(1) actions. This means that the
moment map will be a map
\begin{equation}
\mu:{\cal M}\mapsto \R^3,
\end{equation}
and the manifold ${\cal N}$ will have four fewer dimensions than ${\cal
M}$.
 
The purpose of this Section is to use the \hK quotient construction to
construct $\M$ and to derive some of its properties. Three U(1) actions
are introduced below. Starting from a sixteen-dimensional
manifold, described below, the uncentred $(2,1)$ moduli space will be
obtained by a U(1) \hK quotient. The centre of mass will then be fixed
by a further U(1) \hK quotient, leaving an eight-dimensional space.
The final U(1) quotient will fix the position of the \one-monopole,
leaving a four-dimensional space. If it is fixed at the origin, this
space is the Atiyah-Hitchin manifold. Physically, this corresponds to
taking the limit of large \one-monopole mass. 

The U(1) actions are subgroups of the torus group whose elements are
represented by
\begin{eqnarray}
G&=&\exp{\left[i\frac{s-s_1}{s_2-s_1}\theta_1{\bf 1}_2-i\frac{s-s_1}{s_2-s_1}\theta_2\tau_3\right]},\\
g&=&\exp{\left\{i(\theta_1+\theta_2)+i\frac{s-s_2}{s_3-s_2}[\theta_3-(\theta_1+\theta_2)]\right\}}.\nonumber
\end{eqnarray}
Roughly speaking, this U(1)$^3$ contains a gauge U(1), the U(1) of
overall phase and a U(1) of relative phase. The gauge U(1) has
$\theta_1=\theta_2$ and $\theta_3=0$.

We want to examine these U(1) actions on the moduli
spaces. Coordinates on the orbit of the group are given by 
\begin{eqnarray}
\phi_1&=&i\int_{s_1}^{s_2}\tr{{T}_0}ds,\\
\phi_2&=&-i\int_{s_1}^{s_2}\tr({{T}_0\tau_3})ds,\nonumber\\
\phi_3&=&i\int_{s_2}^{s_3}{t}_0ds\nonumber
\end{eqnarray}
and these transform as
\begin{eqnarray}
\phi_1&\rightarrow&\phi_1+2\theta_1,\\
\phi_2&\rightarrow&\phi_2+2\theta_2,\nonumber\\
\phi_3&\rightarrow&\phi_3+\theta_3-\theta_1-\theta_2.\nonumber
\end{eqnarray}
In order to specify a U(1) subgroup of this U(1)$^3$ we write for example
\begin{eqnarray}
\theta_1&=&\alpha \theta,\\
\theta_2&=&\beta \theta,\nonumber\\
\theta_3-(\theta_1+\theta_2)&=&\gamma \theta \nonumber
\end{eqnarray}
and then specify a $\RP^2$ vector $(\alpha,\beta,\gamma)$. The gauge
transformation is given by $(1,1,-2)$. Let us write ${\cal
  S}(\alpha,\beta,\gamma)$ for the U(1) specified by $(\alpha,\beta,\gamma)$.
The moment map for  ${\cal S}(\alpha,\beta,\gamma)$ is a map
\begin{equation} 
\mu(\alpha,\beta,\gamma): {\cal M}\rightarrow\R^3,
\end{equation}
and is
\be
\mu_i(\T;\alpha,\beta,\gamma)=\alpha\tr{{T}_i} -\beta
\tr({T}_i \tau_3)+\gamma {t}_i.\label{mm}
\ee

Of course, the moment map only serves to define the level set. To
calculate the \hK manifold the level set must be quotiented by the
U(1) action. The quotient by ${\cal S}(\alpha,\beta,\gamma)$ is
performed, in effect, by introducing a group action fixing condition
of the form
\begin{equation}
  A\phi_1+B\phi_2+C\phi_3=0.
\end{equation}
This fixing condition must be invariant under group transformations
whose Killing vectors are orthogonal to the ${\cal
  S}(\alpha,\beta,\gamma)$ Killing vector
$\Y(\alpha,\beta,\gamma)$ which is
\begin{equation}
\left( \frac{\alpha i}{M}{\bf 1}_2-\frac{\beta i}{M}\tau_3,{\bf 0}_2, {\bf
0}_2, {\bf 0}_2\right),
\end{equation}
on the lefthand interval, with ${\bf 0}_2$ the zero matrix, and
\begin{equation}
\left( \frac{\gamma i}{m},0,0,0 \right),
\end{equation}
on the righthand interval. In the obvious notation
\begin{equation}
(\Y,\Y')=\frac{2}{M}(\alpha\alpha'+\beta\beta')+\frac{1}{m}\gamma\gamma'.
\end{equation}
Furthermore,
\begin{equation}
\delta(A\phi_1+B\phi_2-C\phi_3)= -2\alpha' A-2\beta'B-\gamma'C
\end{equation}
under ${\cal S}(\alpha',\beta',\gamma')$.  Thus 
\begin{equation}\label{ABC}
(A,B,C)=\left(\frac{\alpha}{M},\frac{\beta}{M},\frac{\gamma}{m}\right)
\end{equation}
is invariant under group transformations orthogonal to ${\cal
  S}(\alpha,\beta,\gamma)$. In terms of Nahm data the quotient condition is
\begin{equation}
  \frac{\alpha}{M} \int_{s_1}^{s_2} \tr{{T}_0}ds -\frac{\beta}{M}
    \int_{s_1}^{s_2} \tr({{T}_0\tau_3})ds +\frac{\gamma}{m}
    \int_{s_2}^{s_3}{t}_0ds=0.
\end{equation}
This complements the moment map conditions.

\subsection{HyperK\"ahler quotient construction of $\M$}
\label{HKQsect}

\ns The moment map of the gauge action on $\M$ is 
\begin{equation}
\mu_i(\T;1,1,-2)=\tr{{T}_i} - \tr({T}_i \tau_3)-2 {t}_i. \label{mmg}
\end{equation}
and so the level set of the gauge transformation moment map
$\mu^{-1}({\bf 0};1,1,-2)$ satisfies the fixing condition ${t}_i
=({T}_i(s_2))_{2,2}$. What this means is that the moduli space $\M$ is
the \hK quotient of the space of unmatched Nahm data by
this part of the gauge transformation. In this Section, we use this \hK
quotient to attach a \one-monopole to a \two-monopole. This gives an
alternative derivation of the metric on $\M$.

\np We have seen how the lefthand Nahm data are the same, whether or not
$m=0$ and that in this sense the space of $(2,[1])$-monopoles can be
interpreted as the space of \two-monopoles. We now consider
the sixteen-dimensional manifold which will form the starting point for
our series of U(1) \hK quotients. This is the direct product of the
space of uncentred  $(2,[1])$-monopoles, with the moduli space of a
1-monopole. The moduli space of a 1-monopole is $\R^3 \times
\mbox{S}^1$. This describes the position and phase of the monopole. The
Nahm data for this product are identical to the Nahm data for the space
of SU(5) $(2,[1],[0],1)$-monopoles depicted below 
\begin{equation}
\begin{array}{c}
\begin{picture}(295,65)(55,615)
\thinlines
\put( 60,620){\line( 1, 0){214}}
\put(100,620){\line( 0, 1){ 40}}
\put(100,660){\line( 1, 0){ 60}}
\put(160,660){\line( 0,-1){ 20}}
\put(160,640){\line( 1, 0){  2}}
\put(162,640){\line( 0,-1){ 20}}
\multiput(160,640)(0.00000,-8.00000){3}{\line( 0,-1){  4.000}}
\put(164,620){\line( 0, 1){20}}
\put(164,640){\line( 1, 0){80}}
\put(244,620){\line( 0, 1){20}}
\put(162,620){\line( 0,-1){  5}}  
\put(100,620){\line( 0,-1){  5}}
\put(160,620){\line( 0,-1){  5}}
\put(164,620){\line( 0,-1){  5}}
\put(244,620){\line( 0,-1){  5}}
\put(84,638){\makebox(0,0)[lb]{\smash{\SetFigFont{12}{14.4}{rm}2}}}
\put(95,623){\vector(0,1){  36}}
\put(95,623){\vector(0,-1){  0}} 
\put(254,626){\makebox(0,0)[lb]{\smash{\SetFigFont{12}{14.4}{rm}1}}}
\put(249,623){\vector(0,1){  15}}
\put(249,623){\vector(0,-1){  0}} 
\put( 97,605){\makebox(0,0)[lb]{\smash{\SetFigFont{12}{14.4}{rm}$s_1$}}}
\put(159,605){\makebox(0,0)[lb]{\smash{\SetFigFont{12}{14.4}{rm}$s_2$}}}
\put(241,605){\makebox(0,0)[lb]{\smash{\SetFigFont{12}{14.4}{rm}$s_3$}}}
\end{picture}\end{array}
\label{su5diagram}
\end{equation}
where, as before, $M=s_2-s_1$ and $m=s_3-s_2$. There is a gap between the
lefthand and righthand Nahm data to represent their independence from each
other. In the SU(5) analogue, the gap represents the
$(\;,\;,[0],\;)$ part of the charge.

\np The $(2,[1],\;,\;)$-monopoles do not interact with the
$(\;,\;,\;,1)$-monopole. The sixteen dimensions of the moduli
space of these monopoles separate into twelve moduli parameterising the
space of uncentred $(2,[1])$-monopoles and four parameterising the
position and phase of the 1-monopole.  The Nahm data $({T}_0, {T}_1,
{T}_2, {T}_3)$ at
$s=s_2$ are $\mathfrak{u}(2)$-valued and the Nahm data
$\left({t}_0, {t}_1, {t}_2, {t}_3 \right)$ on
the interval $[s_2,s_3]$ are a quadruplet of complex numbers. The metric
on this space is 
\begin{equation}
  ds^2 = ds^2_{(2,[1])} + m d {\bf r}^2 + m d r_0^2,\label{12dmet}
\end{equation}
where ${\bf r}$ is the position of the 1-monopole given by ${\bf
  r}=-i{\bf {t}}$ and $r_0 = \frac{i}{m}\int_{s_2}^{s_3}
{}t_0ds$. 
 
The hyperK\"ahler quotient of the sixteen-dimensional metric
(\ref{12dmet}) by ${\cal S}(1,1,-2)$ is now performed. The three
moment map equations are 
\begin{equation}
  {t}_i=(T_i(s_2))_{2,2}.
\end{equation}
These are precisely what is required; they identified the position of
the \one-monopole with a point on the nonAbelian cloud of the
$(2,[1])$-monopole. What remains is the calculation of the
$(2,1)$-metric without including the interaction of the phases of the
\two-monopole and \one-monopole.

To complete the hyperK\"ahler quotient construction, we must project the
tangent vectors to be orthogonal to the tangent vector which generates
the U(1) action. This amounts to removing from the metric the one-form
which is preserved by the U$(1)$ action. This is achieved by setting
\begin{equation}
  \frac{1}{M}(\phi_1+\phi_2)-\frac{2}{m}\phi_3=0.
  \label{phasematch}
\end{equation}
The first two terms above combine into the one-form corresponding to change of
phase of the nonAbelian cloud of the $(2,[1])$ monopole. The third term
is the phase of the \one-monopole. Thus the U(1)-fixing condition
(\ref{phasematch}) has a physical interpretation; it can be thought of as
matching the phase angle of the cloud in the Dancer monopole with the
phase angle of the \one-monopole. Imposing this condition on the metric
is equivalent to requiring that any tangent vector in the metric is
orthogonal to the Killing vector ${\cal Y}(1,1,-2)$. Performing the
hyperK\"ahler quotient explicitly yields the metric (\ref{fullmetric}).

\subsection{Fixing the centre of mass using a \hK quotient}\label{hKCoM}

\ns In this Sectionm we consider the torus action on the uncentred
moduli space $\M$ derived above. The Nahm data are matched but not
centred and so the moment map (\ref{mm}) is
\begin{equation}
  \mu_i(\T;\alpha,\beta,\gamma)=(\alpha-\beta) \tr{{T}_i}+
    (\gamma+2\beta) {t}_i.
\end{equation}
Thereforem the centre of mass is fixed by a U(1) action with
\begin{equation}
  \frac{\gamma+2\beta}{\alpha-\beta}=\frac{m}{M}.
\end{equation} 
Requiring that this action is orthogonal to ${\cal S}(1,1,-2)$ implies 
\begin{equation}
  \gamma=\frac{m}{M}(\alpha+\beta)
\end{equation} 
and hence the moment map is given by
$(\alpha,\beta,\gamma)=(M,0,m)$. This means that the overall
phase is set to zero, that is $(A,B,C)$=$(1,0,1)$. This defines $\M^0$ as a
\hK quotient of $\M$. It also demonstrates that the overall phase is
$\phi_1+\phi_3$. 

Equally well, the torus action could have been used to fix the traces
of the lefthand Nahm data. The moment map is $\mu_i(\T)\propto
{\rm trace}{T_i}$ if $\gamma+2\beta=0$. Thus, the U(1) action which sets
$\tr{T_i}=0$ and is orthogonal to gauge transformations is given by
\begin{equation}
  (\alpha,\beta,\gamma)= \left(-\frac{2M+m}{m}, 1, -2\right).
\end{equation}
The fixing condition in this case is given by $(A,B,C)=\left( -(2M+m)/m,1,
-2M/m \right)$. This seems at first sight to be unusual;
the group action fixing condition is clarified by writing it as
\begin{equation}
  -\frac{2M+m}{m}(\phi_1+\phi_3) +(\phi_2+\phi_3) =0.
  \label{fixcond}
\end{equation} 
We see that the fixing condition is the sum of two gauge-invariant
pieces. The first is the total phase of the $(2,1)$-monopole system
and the second is the relative phases of the \two-monopole and the
\one-monopole. 

The residual U(1), orthogonal both to the gauge
U(1) and ${\cal S}(-(2M+m)/m,1,-2)$, is ${\cal S}(0,M,m)$. By
(\ref{ABC}) a coordinate on this U(1) is $\phi_2+\phi_3$. This
coordinate is the coordinate $\alpha_8$ adopted earlier. This can be
verified by calculating $\phi_2$ explicitly in terms of the
group action parameters.

In fact, any point along the line joining $\tr{T_i}$ and $t_i$ can be
fixed. However, all these manifolds are isometric up to a change in
the \one-monopole mass. The proof of this is essentially the same as
the demonstration in Section \ref{CoMs} that the $\M^0$ metric can be
calculated from traceless Nahm data.

\subsection{A \hK quotient of $\M^0$}\label{hKfm}

\ns In \cite{D}, the \hK quotient construction is used to
construct a one-parameter family of four-dimensional \hK
manifolds from the moduli space $\MD$.  In \cite{H}, it was noted that
these manifolds can be thought of as an infinite mass limit in which
the \one-monopole in $\M$ becomes fixed in position. This is because
the moment map fixes the \one-monopole degrees of freedom. In this
Section, a similar \hK quotient is performed on $\M$. It is
found that the resulting four-dimensional manifolds are the same as
those derived from $\MD$.

\np $\M^0$ has an isometric action of U(1) given
by ${\cal S}(1,1,2m/M)$. The \hK quotient
of $\M^0$ results in a four-dimensional
\hK manifold 
\begin{equation}
{\cal M}^{4}({\bf y};m)=\mubf^{-1}({\bf y},1,1,2m/M)/\mbox{U}(1),
\end{equation}
where
${\bf y}\in{\R}^3$. The mass of the
\one-monopole is a parameter for the moduli space 
${\cal M}^{0}_{(2,1)}$  and so it might be expected that
it also parameterises ${\cal M}^{4}({\bf y};m)$. For convenience,
this possible dependence on $m$ has been denoted explicitly. In fact,
one of the results of this Section is that ${\cal M}^{4}({\bf y};m)$ is
actually independent of $m$. 

\np The space of $(2,[1])$-monopoles is the $m\rightarrow0$ limit of
the space of $(2,1)$-monopoles and so, in our notation, the
\hK manifolds constructed in \cite{D} are denoted ${\cal
  M}^4({\bf y};0)$.  In fact, $\MD^0$ has a
triholomorphic action of SU(2) and any U(1) subgroup of SU(2) can be
used to perform the \hK quotient. This appears to give a
larger class of quotient spaces than there are for the $(2,1)$ case.
However, this is not the case because every U(1) subgroup results in the
same quotient space.

\np Although $\bf{y}$ is an $\R^3$ vector, there is only a
one-dimensional space of ${\cal M}^4({\bf y};m)$ for given $m$.
This is because the action of SO(3) on $\MM{0}$ maps the space ${\cal
M}^4({\bf y};m)$ isometrically to the space ${\cal M}^4(R{\bf y};m)$,
where $R\inn$SO(3). This means that the three-parameter dependence of ${\cal
M}^4({\bf y};m)$ on ${\bf y}$ reduces to a one-parameter dependence on
$|{\bf y}|$. 

\np Since the moment map for the ${\cal G}^{0,0,\star}/{\cal G}^{0,0,0}$ action
on $\MM{0}$
is
\begin{equation}
\mu_i(\T)\propto t_i
\end{equation}
the moment map fixes the position of
the \one-monopole, just as in the minimal symmetry breaking case.  

\np The spaces ${\cal M}^4({\bf y};m)$ have an SO(2) isometry
corresponding to the SO(2) subgroup of SO(3), which fixes the vector
${\bf y}$. If ${\bf y}={\bf 0}$, the whole SO(3) acts as an isometry.
Since we have an expression for the metric on $\M$, it is possible to
explicitly perform the quotient and derive the manifold ${\cal M}^4
({\bf 0};m)$.  The constraint ${\bf y}={\bf 0}$ implies the $t_i$ are
all zero. Substituting this into the full metric (\ref{fullmetric})
gives the metric on the level set $\mubf^{-1}({\bf 0})$. Explicitly,
$t_i=0$ is equivalent to
\begin{eqnarray}
D&=&\frac{1}{M}K(k),\\
\theta&=&\pi/2,\nonumber\\
\chi&=&0,\nonumber
\end{eqnarray}
since $\cn{k}{K(k)}=0$. Here and below, we will use the standard complete
elliptic integrals $K\equiv K(k)$ and $E\equiv E(k)$. Substituting this
into (\ref{fullmetric}) gives  
\begin{equation}
  ds^2=\frac{1}{4}\left(\frac{b^2}{K^2}dK^2+a^2\sigma_1^2+b^2\sigma_2^2+
    c^2\sigma_3^2\right)+\frac{1}{2}\frac{EK}{M+mEK}\left
    (d\phi-\frac{k'K}{E}\sigma_1\right)^2,
\label{U1bundle}
\end{equation}
where  
\begin{eqnarray}\label{AHfxns}
  a^2&=&\frac{2}{M}\frac{K(K-E)(E-k'^2K)}{E},\\
  b^2&=&\frac{2}{M}\frac{EK(K-E)}{E-k'^2K},\nonumber\\
  c^2&=&\frac{2}{M}\frac{EK(E-k'^2K)}{K-E}.\nonumber
\end{eqnarray}

\np Now, to derive ${\cal M}^4({\bf 0};m)$ itself, the U(1) action
must be quotiented out. This action corresponds to 
\begin{equation} 
\phi\rightarrow\phi+\phi_0,
\end{equation}
and the U(1) is quotiented out by discarding the
$(d\phi-k'K\sigma_1/E)^2$ term in (\ref{U1bundle}). The resulting
metric is, 
\be\label{AH}
ds^2=\frac{1}{4}\left(\frac{b^2}{K^2}dK^2+a^2\sigma_1^2+b^2\sigma_2^2+
c^2\sigma_3^2\right),
\end{equation}
which is the Atiyah-Hitchin metric up to a scale of a quarter.

\np In \cite{D}, it was shown that when ${\bf y}={\bf 0}$ and $m=0$
the \hK quotient gives the double cover of the
Atiyah-Hitchin manifold. From the above, we see that the metric on
${\cal M}^4 ({\bf 0};m)$ is independent of $m$ and ${\cal M}^4 ({\bf
  0};m)$ is the double cover of the Atiyah-Hitchin manifold for all
values of $m$. This is not surprising as ${\cal M}^4({\bf 0};m)$ is a
four-dimensional \hK manifold with an isometric SO(3) action
and there is only a small number of manifolds with these properties.
Furthermore, ${\cal M}^4 ({\bf 0};m)$ is the double cover of the
Atiyah-Hitchin manifold when $m=0$. Therefore, if $m$ is varied, the only
possible variation of the metric is by an overall scaling. This point
has also been made in \cite{LL}.

\np The manifold $\mubf^{-1}({\bf 0})$ is a U(1) bundle over the
double cover of the Atiyah-Hitchin manifold.  The metric on
$\mubf^{-1}({\bf 0})$ induced from the metric on $\MM{0}$ defines a
U(1) connection on the bundle. The pullback of the curvature of the
U(1) connection to the double cover of the Atiyah-Hitchin manifold is
a closed two-form. In \cite{D2}, this two-form was shown to be SO(3)
invariant and anti-self-dual. It goes to minus itself under the
transformation
\begin{equation}\label{sigma1}
(\sigma_1,\,\sigma_2,\,\sigma_3)\rightarrow(-\sigma_1,\,-\sigma_2,\,
\sigma_3).
\end{equation} 
Thus, it is the well known Sen form \cite{GRu, S}.  This can be seen
directly from (\ref{U1bundle}) since the connection is $A=H\sigma_1$
with $H=-k'K/E\,$.  The resulting curvature $F=dA$ is
\begin{equation}
F=Hd\sigma_1+\frac{dH}{dK}dK\wedge\sigma_1,
\end{equation} 
and this is the Sen form.

\np It does not seem possible to perform the \hK quotient
tractably for ${\bf y}\neq {\bf 0}$. However, it can be shown that the
resulting manifold does not depend on $m$. From the Nahm construction,
the only $m$ dependence in the metric arises from the terms
\begin{equation}
  m\sum_\mu x_\mu z_\mu,
\end{equation}  
in the inner product (\ref{innerpr}) of tangent vectors ${\cal
  X}$ and ${\cal Z}$. Since the moment map fixes $t_i$, this means that
tangent vectors on $\mubf^{-1}({\bf y})$, given by ${\cal
  X}_{\mu}=(X_{\mu},x_{\mu})$ have $x_i=0$. Performing the U(1)
quotient on $\mubf^{-1}({\bf y})$ to give ${\cal M}^4({\bf
    y};m)$ amounts to projecting tangent vectors so that they are
  orthogonal  to
the tangent vector that generates this U(1) action. The tangent
vector which generates this action has $x_0$ as its only nonzero
component. This means that the orthogonal projection sets the
$x_0$ component of that tangent vector to zero. Thus, there is no
$m$ dependence in ${\cal M}^4({\bf y};m)$.  This supports the
conjecture that there is only a one-parameter family of \hK
deformations of the Atiyah-Hitchin manifold.

\section{Some properties of $\M$}
\label{propssect}
\news
\ns In this Section, we investigate some of the properties of the $(2,1)$
metric. In Section \ref{ratmapsubsec},
the rational map description of the moduli space is used to derive the
topology of $\M$ and of $\M^0$. In Section \ref{submansect}, the metric
on the geodesic submanifold of axially symmetric $(2,1)$-monopoles is
derived. By examining the behaviour of the monopoles on this geodesic
submanifold, it is possible to infer something of $(2,1)$-monopole
dynamics.

\subsection{Rational maps}
\label{ratmapsubsec}

\ns In this Section, the rational map description of $\M$ is
discussed. The space $\M$ is diffeomorphic to the space
of based rational maps from ${\C}$ to the space of total flags in
${\C}^3$, denoted \FC. A total flag inside an $n$-dimensional vector space
$V_n$ is a series of vector subspaces $0\subset V_1\subset
V_2\subset\ldots\subset V_{n-1}\subset V_n$, where $V_i$ has dimension
$i$. Thus, an element of \FC\/ is a pair consisting of a complex plane
and a complex line lying in that plane.  A rational map from \C\/ to \FC\/
consists of a holomorphic map from \C\/ to a line in $\C^3$ and another
holomorphic map from $\C$ to the plane in $\C^3$ containing the
aforementioned line. These maps are not independent, since the image
line of a point in $\C$ must lie in the image plane. However, each of
the maps has a separate degree, these degrees are the topological
charges of the corresponding SU(3) monopole.

\np This diffeomorphism was originally described as a diffeomorphism
between the rational map space and the moduli space of Nahm data \cite{Hu}. 
A more recent and more direct description of the diffeomorphism uses
the Hitchin equation to construct the rational map directly from the monopole
field \cite{J1}.  

\np The Hitchin equation is a scattering  equation along a fixed oriented
line $v$ in ${\R}^3$, 
\begin{equation}
(D_v-i\Phi)u=0.
\label{Scatter}
\end{equation}
$D_v$ is the covariant derivative in the $v$ direction. If we choose
$v$ to be in the $x_3$ direction then there is a complex plane of
lines parameterised by $z=x_1+ix_2\in{\C}$. The
rational map is obtained by considering the solutions of the Hitchin
equation (\ref{Scatter}) along these lines. 

\np The Hitchin equation has three independent
solutions. The asymptotic behaviour of $\Phi$ is known
and substituting from (\ref{1overrF}) shows 
there is a spanning set of solutions $u_1$, $u_2$ and $u_3$ with the behaviour
\begin{eqnarray}\label{Scatter1}
\lim_{x_3\rightarrow\infty}u_1(x_3;z)x_3^{-k_1/2}e^{s_1x_3}&=& e_1,\\
\lim_{x_3\rightarrow\infty}u_2(x_3;z)x_3^{(k_1-k_2)/2}e^{s_2x_3}&=&e_2,\nonumber\\
\lim_{x_3\rightarrow\infty}u_3(x_3;z)x_3^{k_2/2}e^{s_3x_3}&=& e_3,\nonumber
\end{eqnarray}
where the $e_i$ are the unit eigenvectors of 
${\displaystyle \lim_{x_3\rightarrow\infty}{\Phi}}$. 

\np Thus, the $x_3\rightarrow\infty$ asymptotic Hitchin equation
nominates three particular solutions distinguished by their rate of
growth or decay. The $x_3\rightarrow-\infty$ asymptotic Hitchin
equation also distinguishes three solutions. These are used to
construct the flag.  In the three-dimensional space of solutions, there
is a one-dimensional subspace generated by the solution which decays at
the fastest rate as $x_3\rightarrow-\infty$ and a two-dimensional
subspace spanned by this solution and the next fastest decaying
solution. These subspaces can be written as linear combinations of the
$u_1$, $u_2$ and $u_3$ and the linear coefficients define a total flag
in $\C^3$, that is, an element of \FC. The flag depends on $z$ and the
Bogomolny equations then imply that this flag varies holomorphically in
$z$. The monopole charge determines the degree of the map. The monopole
boundary conditions for large $z$ show that the map is based. This means
that as $z\rightarrow\infty$ the map approaches a fixed element in \FC.

\np Thus, we are interested in based rational maps of degree $(2,1)$. It
is not difficult to write down the most general map of this type. Rather
than describing the plane by a pair of holomorphic lines spanning it, we
describe it by specifying a line in the plane and a line perpendicular
to it. This perpendicular line is antiholomorphic. Thus, the general
degree $(2,1)$ based rational map is

\begin{eqnarray}
  \label{Ratmap}
  E_1&=&\left(1,\frac{az+b}{z^2+cz+d},\,\frac{ez+f}{z^2+cz+d}\right),\\
  E_2&=&\left(\frac{\alpha}{z-\gamma},\,\frac{\beta}{z-\gamma},1\right),
    \nonumber
\end{eqnarray} 
where $E_1$ describes a line in ${\C}^3$ and $E_2^\star$
describes the line perpendicular to a plane in ${\C}^3$.
$E_2^\star$ is the complex conjugate of $E_2$. For
$E_1$ to lie in the plane defined by $E_2^\star$ the Hermitian inner
product of $E_1$ and $E_2^\star$ must vanish. 

\np If $af-be=0$, then it follows from the orthogonality condition that the
map is not genuinely of degree $(2,1)$. Either $E_1$ is not a genuine
degree two map, because $z=-b/a$ is a root of $z^2+cz+d$ as well as of
$az+b$ and $ez+f$, or $E_2$ is degree zero because $\alpha=\beta=0$.
This means that $af-be\neq 0$ and orthogonality determines $\alpha$,
$\beta$ and $\gamma$ in terms of $a$, $b$, $c$, $d$, $e$ and $f$. In
other words, $E_2$ is determined from $E_1$. This means that the
space of degree $(2,1)$ rational maps is isomorphic to
$(a,b,c,d,e,f)\in{\C}^6$ with the constraint $af-be\neq 0$.  This is
$\mbox{GL}(2,{\C})\times{\C}^2$. Thus, the moduli space is
topologically equivalent to $\mbox{GL}(2,{\C})\times{\C}^2$ or,
equivalently, U(2)$\times{\R}^8$.

\np The manifold $\M$ has the product form
\begin{equation}\label{m21}
\M=
{\R}^3\times\frac{{\R}\times\MM{0}}{\Z}.
\end{equation}
$\MM{0}$ is the relative moduli space obtained
by quotienting out the centre of mass action, ${\R}^3\times\,\R$.
Using (\ref{Scatter}), it is possible to determine the effect of
translations and gauge transforms on the map.  A translation in the
$(x_1,x_2)$ plane by $w\in{\C}$ acts as
\begin{eqnarray}
E_1(z)&\rightarrow& E_1(z-w),\\
E_2(z)&\rightarrow& E_2(z-w).\nonumber
\end{eqnarray}
This action can be used to set $c=0$. 

\np A translation in the $x_3$ direction by $\lambda$ and a gauge
transform $g=e^{\theta \Phi}$ has the following effect: 
\begin{equation}
E_1(z)\rightarrow\left(1,\,\frac{e^{M(\lambda+i\theta)}(az+b)}
  {z^2+cz+d},\,\frac{e^{(M+m)(\lambda+i\theta)}(ez+f)}{z^2+cz+d}\right).
\end{equation} 
The transformation on $E_2$ is determined by the transformation on
$E_1$.  The gauge transformation $g$ changes the overall phase
of the monopole. Notice that $\theta$ is an angle only when $m/M$ is
rational. Under this transformation 
\begin{equation} 
af-be \rightarrow e^{(2M+m)(\lambda+i\theta)}(af-be),
\end{equation}
and so $af-be$ can be set to one. Setting
$c$ to zero and $af-be$ to one fixes the centre of mass and overall
phase, allowing us to conclude $\MM{0}$ is
determined by $(a,b,d,e,f)\in{\C}^5$ such that $af-be=1$.  This
implies that $\MM{0}$ is topologically equivalent to
SU(2)$\times{\R}^5$. The $\Z$ quotient in the product form (\ref{m21})
follows from the identification
\begin{eqnarray}
\theta &=& \theta+\frac{2\pi n}{2M+m},\\
(a\, ,\,b) &=& e^{-2\pi ni\frac{M}{2M+m}}\,(a\,
,\,b), \nonumber\\
(e\, ,\,f) &=& e^{-2\pi ni\frac{M+m}{2M+m}}\;(e\,
,\,f),\nonumber
\end{eqnarray}
where $n\in\Z$.

\np We remark that this agrees with the results in \cite{D, D2} for
the minimal symmetry breaking case. This is to be expected since
$\MD^0$ is the smooth $m\rightarrow 0$ limit of $\MM{0}$ and the
topological properties should be identical, as indeed they are. The
manifold is given in this case by
\begin{equation}
\MD={\R}^3\times\frac{S^1\times{\cal M}^0_{(2,[1])}}{{\Z}_2},
\end{equation}
where ${\cal M}^0_{2,[1]}$ is the moduli space of centred
$(2,[1])$-monopoles.  Since $m=0$ here, the total phase is periodic
and the ${\Z}$ quotient reduces to a ${\Z}_2$ quotient. Here, $\MD$ is
the double cover of what is denoted $M^8$ in \cite{D}.  The rational
map description of $\MD$ is given by maps from $\C$ to $\CP^2$, whose
images do not lie in a $\CP^1$. This means that the rational map is given
by $E_1$. $E_1$ does not lie in a $\CP^1$ if $af-be\neq 0$ and so this
constraint applies for $(2,[1])$, as well as for $(2,1)$.  In case of
$(2,1)$, we have $E_2$ in addition to $E_1$ but $E_2$ is determined by
$E_1$.

\subsection{Geodesic submanifolds}
\label{submansect}

\ns The analysis of geodesics on monopole moduli spaces is a formidable
task. However, the existence of tractable geodesic submanifolds allows
one to partially deduce their behaviour. In this Section, we locate
the geodesic submanifold of axially symmetric $(2,1)$-monopoles
Since the metric on $\MM{0}$ is now known, we can find the induced
metric on the submanifold and use this to study the behaviour of
axially symmetric monopoles.

\np The fixed point set of a isometric action on a Riemannian manifold
is a totally geodesic submanifold. The geodesics on the submanifold
are also geodesics on the manifold itself. This allows us to study
$(2,1)$ geodesic behaviour using symmetry.

\np There is an isometric SO(3) action on $\MM{0}$ so the fixed point
set of any subgroup of SO(3) will be a geodesic submanifold of
$\MM{0}$ consisting of monopoles invariant under the subgroup. A
compensating U(1) transform in ${\cal G}^{0,0,\star}/{\cal G}^{0,0,0}$
may be needed to keep the monopole invariant. This action is also
isometric and if the combined transformations are a subgroup of
SO(3)$\times$U(1), then the fixed point set will still be a geodesic
submanifold.  The SO(3) and U(1) actions are defined on the the Nahm
data. Symmetric Nahm data need not be strictly invariant under the
relevant group action; it will generally change by a gauge
transformation. To be precise about this, we need
to check that the coordinates that describe a point on $\MM{0}$ are
invariant under the group action. Such points, if any, then comprise
the geodesic submanifold.  Given a subgroup of SO(3), we look for Nahm
data such that the eight coordinates: $\alpha_a$, $a=1\ldots8$ 
of Section \ref{1formsect} are invariant under a
combined subgroup of SO(3)$\times$U(1). This will define a geodesic
submanifold.

\subsubsection{Spherical symmetry}

\ns The only lefthand Nahm data $T_i$ invariant under the SO(3) action,
(\ref{SO3action}), has the form 
\begin{equation}
T_i=-\frac{1}{(s-s_1)}e_i,  
\end{equation}
that is $D=k=0$.  From (\ref{ti}) the \one-monopole will be
positioned at 
\begin{equation}
-it_i=\frac{1}{2}(0\,,\,0\,,-1/M)_i.
\end{equation}
However, spherically symmetric Nahm data must have $t_i=0$. This means
that there is no spherically symmetric $(2,1)$-monopole.  In the
$m\rightarrow 0$ limit, a spherically symmetric monopole appears. This
has been known for some time \cite{BW}.

\subsubsection{Axial symmetry}

\ns Monopoles which are axially symmetric about the $x_3$-axis can be
obtained by requiring $f_1(s)=f_2(s)$ in the ansatz (\ref{basicdata}).  We
must also require $t_1=t_2=0$.  This is achieved if $k=1$ and $\theta$
is set to zero or $\pi$ in (\ref{ti}). Setting $k=1$ in the Euler top
functions gives
\begin{eqnarray}
  f_1(s)=f_2(s)&=&-D\mbox{cosech}\left( D(s-s_1)\right),\\
  f_3(s)&=&-D\mbox{coth}\left( D(s-s_1)\right),\nonumber
\end{eqnarray}
where, for convenience, we have set $M=s_2-s_1=1$.  The \one-monopole is
positioned at $(0,\,0,\,\mp\frac{1}{2}D\coth D)$ with the sign depending
on whether $\theta$ is zero or $\pi$. This is very similar to the
axially symmetric Nahm data constructed in \cite{D}. The $(2,[1])$
Nahm data are axially symmetric, if the action in (\ref{SO3action})
with $G(s_1)=e^{i\alpha \tau_3/2}$ is combined with an action $G'\in{\cal
  G}^{0,\star}$ that has $G'(s_2)=e^{i\alpha\tau_3/2}$. This means
that a
transformation of this form will leave $\alpha_1$ to $\alpha_7$
unchanged since these are coordinates when $m=0$. Requiring the
relative phase (\ref{relphase}) to be unchanged determines what the
compensating ${\cal G}^{0,0,\star}/{\cal G}^{0,0,0}$ transformation
must be.  In fact, it is easy to see by inspection that ${\alpha}_8$
will be invariant if $t_0$ is unchanged by the combined
transformation. This implies that $g(s_3)=e^{-i\alpha/2}$ since
$\left( G'(s_2)\right)_{2,2}=e^{-i\alpha/2}$.  Thus the $(2,1)$ Nahm
data are invariant under a combined transform of (\ref{SO3action}) and
${\cal G}^{0,0,\star}$ with $G(s_1)=e^{i\alpha\tau_3/2}$ and
$g(s_3)=e^{-i\alpha/2}$. The combined transform is thus a diagonal U(1)
subgroup of SO(3)$\times$U(1) and the Nahm data are fixed under this
action. $D$ can take any positive value. 

\np These Nahm data can be interpreted in terms of monopole
configurations. The \one-monopole position is $-it_i$. We will assume
that the energy density of the \two-monopole configuration does not
change significantly with changes of $m$, the mass of the
\one-monopole. This means that we assume the configuration is similar to
the corresponding configuration of $(2,[1])$-monopoles.
Thus, we follow \cite{DL2} and interpret $D$ as the separation of the
\two-monopoles (in the asymptotic metric calculations of Section
\ref{ptdymet}, $D$  is used with success as a separation parameter). We
also assume $k=0$ represents toroidal \two-monopoles \cite{DL2}. For
large $D$, these Nahm data represent two \onel-monopoles separated along
the $x_3$-axis with approximate positions
$(0,\,0,\,\pm\frac{1}{2}D)$. Since the \one-monopole is positioned at
$(0,0,\frac{1}{2}D\coth D)$ or $(0,0,-\frac{1}{2}D\coth D)$ it is close
to one of the \onel-monopoles.  When it is very close, the configuration
should look like a spherically symmetric $(1,1)$-monopole well separated
from a \onel-monopole. 

\np These Nahm data remain axially symmetric if acted on by the U(1)
group ${\cal G}^{0,0,\star}/{\cal G}^{0,0,0}$ whose action commutes with
the SO(3) action. Thus, there is a two-parameter family of axially
symmetric monopoles. It is the U(1) orbit of the one-dimensional
family parameterised by $D\in[0,\infty)$. We call this family the
hyperbolic region because its Nahm data are hyperbolic.  The induced
metric on the hyperbolic region is easily obtained from
(\ref{fullmetric}) by substituting the constraints above:

\begin{equation}
  \label{hypmetric}
  ds^2=\eta_1(D)dD^2+\frac{\zeta_1(D)}{1+m\zeta_1(D)} d\phi^2,
\end{equation}   
where $\eta_1$ and $\zeta_1$ are given by
\begin{eqnarray}
\eta_1(D)&=&\frac{1}{2}(\sinh D\cosh D-D)(D-\tanh D)\frac{\cosh
D}{D\sinh^3 D}\\
&+&\frac{m}{4} \frac{(\sinh D\,\cosh D -D)^2}{\sinh^4 D},\nonumber\\
\zeta_1(D)&=&\frac{1}{2} \frac{D(D-\sinh D\,\cosh D)}
{\cosh D\,\sinh D(\tanh D -D)}.\nonumber
\end{eqnarray}

The hyperbolic region is only part of the geodesic submanifold of axially
symmetric Nahm data. There is also a trigonometric region, in which
\begin{eqnarray}\label{trigdata}
f_1(s)=f_2(s)&=&-D\mbox{cosec}\left( D(s-s_1)\right),\\
       f_3(s)&=&-D\cot\left(D (s-s_1)\right).\nonumber
\end{eqnarray}
By an argument which is identical to the above, these Nahm data are invariant under
the diagonal subgroup of SO(3)$\times$U(1). They correspond
to $k=0$ Euler top functions. However, if $k$ is set to zero in
(\ref{topfunctions}), the Nahm data which result are axially symmetric about
the $x_1$-axis. The solutions (\ref{trigdata}) correspond to a
different ordering of the $f_1$, $f_2$ and $f_3$ and may be obtained from the
ansatz space by rotation. For $t_1=t_2=0$, $\theta$ is again zero or
$\pi$ and the position of the \one-monopole is given by
$(0,0,\pm\frac{1}{2}D\cot D)$. This is the trigonometric
region. $D\in[0,\pi)$ and the two regions are joined at $D=0$, where
the Nahm data are rational and $k$ is not determined.

In the trigonometric region, the \two-monopoles are
coincident and toroidal in shape. The trigonometric 
Nahm data also remain axially symmetric if acted on by the U(1) factor
${\cal G}^{0,0,\star}/{\cal G}^{0,0,0}$. The metric on the trigonometric
region is

\be
  \label{trigmetric}
  ds^2=\eta_2(D)dD^2+\frac{\zeta_2(D)}{1+m\zeta_2(D)}d\phi^2\,
\end{equation}   
with $\eta_2$ and $\zeta_2$ given by
\begin{eqnarray}
  \eta_2(D)&=&\frac{(\sin D\,\cos D -D)^2}{\sin^4 D}\left(
  \frac{1}{2} \frac{\sin D\,\cos D(D-\tan D)}{D\sin D\,\cos D -D^2}+
  \frac{m}{4} \right),\\
  \zeta_2(D)&=&\frac{1}{2} \frac{D(D-\sin D\,\cos D)}
  {\cos D\,\sin D(\tan D-D)}.\nonumber
\end{eqnarray}

The two regions fit together smoothly at $D=0$ and together form a
geodesic submanifold of ${\cal M}^0_{(2,1)}$.  In \rone, 
$\zeta_1$ continually decreases as $D$ increases and approaches $1/2$ as
$D\rightarrow\infty$.  Thus, \rone $\,$ asymptotes to a cylinder of radius
$1/(2+m)$ as $D\rightarrow\infty$. In \rtwo, $\zeta_2$ continually
increases as $D$ increases with $\zeta_2\rightarrow\infty$ as
$D\rightarrow\pi$.  Thus, \rtwo $\,$ asymptotes to a cylinder of radius $1/m$
as $D\rightarrow\pi$. The whole submanifold can be pictured as a
surface which asymptotes at either end to cylinders of radii $1/(2+m)$
and $1/m$ respectively. In the $m=0$ limit it asymptotes to a cone at
one end: this was studied in \cite{DL1}. The surfaces for differing
values of $m$ are depicted in Figure 2.

\begin{figure}[t]
\begin{center}
\leavevmode
\epsfxsize=14cm\epsffile{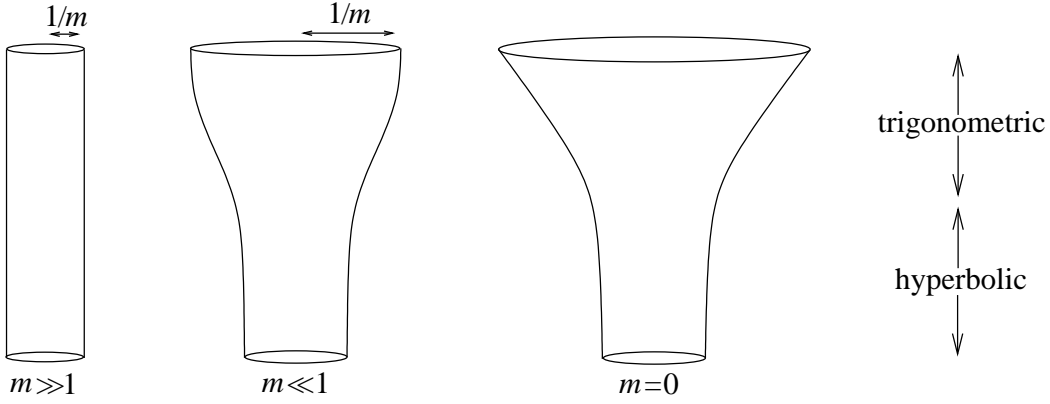}
\caption{The axially symmetric moduli space for differing values of $m$}
\end{center}
\end{figure}

\np Since there is a U(1) isometry, the geodesics are easy to
analyse. The charge 
\begin{equation}
{\cal Q}=\frac{2\zeta\dot{\phi}}{1+m\zeta},
\end{equation} 
is conserved.  Here $\dot{\phi}=d{\phi}/d\tau$, where $\tau$ is a
parameter along the geodesic and we denote by $\zeta$ either $\zeta_1$
or $\zeta_2$ depending on which region the geodesic is in.  Since the
centre of mass and the phase of the \two-monopole are fixed, ${\cal
  Q}$ is the total electric charge of the \one-monopole.

\np The following scattering process occurs. Starting near the
boundary in \rtwo, that is the upper part of the surfaces in Figure 2,
$D$ is close to $\pi$. The \two-monopole is at the origin and its fields
look like those of the axially symmetric embedded SU(2)
two-monopole. The \one-monopole is approaching the \two-monopole from
a large distance along the positive $x_3$-axis.  Since the radius of
the surface is continually decreasing, it is possible that the geodesic
will only travel a certain distance downwards before returning upwards
again. We can rewrite (\ref{hypmetric}) and (\ref{trigmetric}) in the
form
\begin{equation}
  ds^2=\frac{1}{2}dx^2+\frac{\zeta}{1+m\zeta}d\phi^2,
\end{equation}
where
\begin{equation}
  \left(\frac{dx}{dD}\right)^2=2\eta,
\end{equation}
and $x$ increases as the radius of the surface decreases.  There are
two conserved quantities on each geodesic, the electric charge ${\cal Q}$ and
the energy ${\cal E}$. We can write ${\cal E}$ as
\begin{equation}
{\cal E}=\frac{1}{2}\dot{x}^2+\frac{(1+m\zeta)}{4\zeta}{\cal Q}^2.
\end{equation}
The geodesic returns if 
\begin{equation}
m+\frac{1}{\zeta(x_0)}=\frac{4{\cal E}}{{\cal Q}^2}
\end{equation}
for some $x_0$. Holding ${\cal E}$ fixed and increasing ${\cal Q}$
decrease $x_0$, the point where the geodesic returns.

If the electric charge ${\cal Q}=0$, then the geodesic never returns; $x$
keeps increasing and the
\one-monopole passes through the \two-monopole configuration and on to
the negative $x_3$-axis. The geodesic passes from \rtwo\/ into \rone\/ and
the toroidal \two-monopole breaks apart: as $D\rightarrow\infty$ the \one-monopole
is approximately positioned at $(0,0,-D/2)$ and the \two-monopole
configuration becomes particle-like with the \onel-monopoles positioned at
approximately $(0,0,\pm D/2)$. The \one-monopole is asymptotically
coincident with one of the \two-monopoles, giving one 
spherically symmetric $(1,1)$-monopole and one \onel-monopole.

\np It is instructive to compare the ${\cal Q}=0$ geodesic with the
corresponding geodesic in the $\MD^0$. In that case $m=0$, and the
\one-monopole is replaced by a cloud. Starting in the hyperbolic
region, the geodesic describes two widely separated monopoles
approaching each other along the $x_3$-axis. The cloud size is
minimal. The monopole instantaneously forms the spherically symmetric
$(2,[1])$-monopole and then the monopole deforms to a toroidal shape
as it approaches the SU(2) embedded solution. As it does so, the cloud
size continually increases.  This is consistent with the $m\neq 0$
discussion if one adopts the view that the cloud size is a measure of
the distance of the notional massless monopole position from the
position of the massive monopoles. When the \one-monopole is
coincident with one of the \two-monopoles there is no cloud. The cloud
appears only when the \one-monopole is well separated from both of the
\two-monopoles.

\np It is also instructive to compare the behaviour of
$(1,1)$-monopoles with the behaviour of axially symmetric
$(2,1)$-monopoles. In the axially symmetric submanifold a geodesic
with nonzero ${\cal Q}$ may return and so the electric charge gives a
repulsive interaction between the two different types of monopoles.
For large ${\cal Q}$, the \one-monopole approaches the toroidal
\two-monopole configuration but the monopoles slow down and generically they
stop and separate again.  This behaviour agrees with that found in
\cite{C} in the dynamics of $(1,1)$-monopoles. There, the geodesics
were found to be hyperbolae; no bound geodesics exist. The relative
electric charge of the two different types of monopoles has a
repulsive effect.  This leads us to conclude that the interaction of
the two different types of monopoles is generally repulsive.

\section{Asymptotic metrics}
\label{tam}
\news
\ns In this Section, we derive two asymptotic expressions for the
metric. In Section \ref{l1msep}, we consider
the approximate simplification which occurs when the \one-monopole
separation is large. This is useful, because the general behaviour of
the geodesics is very complicated. However, if the monopoles of
different type are repulsive, the generic geodesic will correspond
asymptotically to a \one-monopole well separated from the
\two-monopole configuration. Here the interaction is easy to
understand; the \two-monopoles will interact like SU(2) monopoles but
with a slight modification because of the distant \one-monopole. This
\one-monopole interacts with the \two-monopole in a Taub-NUT like
manner. Indeed, for large separations of the \one-monopole and the
\two-monopole the metric approximately simplifies to a direct product
of the Atiyah-Hitchin metric and the Taub-NUT metric. The corrections
to this simplification are algebraic in the \one-monopole separation
distance.

The asymptotic form of the metric which corresponds to large
separation of the two \onel-monopoles can be
calculated by approximating the monopoles by point dyons using the
methods of \cite{Ma,GM,LWY3}. It has also been calculated by Bielawski
\cite{Bi2} using Nahm data. As with the Taub-NUT approximation to
the Atiyah-Hitchin metric, the point dyon metric has only
exponentially small corrections. At first glance, it seems hard to
imagine how the point dyon metric could be calculated as an
approximation to the $(2,1)$ metric as presented above and so in
Section \ref{ptdymet} we have calculated the radial terms of the
point dyon metric from the $d\alpha_1$ and  $d\alpha_2$
terms of the $(2,1)$ metric. 

\subsection{Large \one-monopole separation}
\label{l1msep}

\ns In this Section, the \one-monopole separation is taken to be
large and the resulting approximate simplification to the metric is
derived. The calculation is a
generalisation of the $(2,[1])$ calculation in \cite{I}; where it is
shown that if the cloud size is large, the $\MD^0$ metric is
approximately the direct product of the Atiyah-Hitchin metric and the
flat $\R^4$ metric.  

\np The \one-monopole separation is large when the $f_i$ are large.
This means that $DM$ must be close to $2K$.  To leading order in
$2K-DM$, $f_1=2r$ and $f_2=f_3=-2r$ where
\begin{equation}
2r\approx \frac{D}{2K-DM}.
\end{equation}
This formula allows us to write $d\alpha_1$ and
$d\alpha_2$ in terms of $dK$ and $dr$.
The incomplete elliptic integrals $g_1$ and $g_2$ can be approximated
by complete integrals giving
\begin{eqnarray}
g_1&\approx&\frac{2}{D^3k^2k'^2}(E-k'^2K)\\
g_2&\approx&\frac{2}{D^3k^2}(K-E).\nonumber
\end{eqnarray}
Furthermore, since $r$ is large $X$ is large and $p_1\approx g_1$, $p_2\approx g_2$ and
$p_3\approx g_1+g_2$ and so $\Omega\approx 1+mr$. Using these formula
working out the approximate metric is just a matter of lengthy calculation
and substitution. The approximate metric is
\begin{equation}
\frac{b^2}{K^2}dK^2+a^2\sigma_1^2+b^2\sigma_2^2+
c^2\sigma_3^2+\left(m+\frac{1}{r}\right)[dr^2+r^2(\hat{\sigma}_1^2+\hat{\sigma}_2^2)]+\frac{r}{1+mr}\hat{\sigma}_3^2
\end{equation}
where the $a^2$, $b^2$ and $c^2$ are the Atiyah-Hitchin functions
defined above (\ref{AHfxns}) and the $\hat{\sigma}_i$ are 
\bea
  \hat{\sigma}_1 &=& d\theta +\sin{\chi}\sigma_1 -\cos{\chi}\sigma_2,
    \\
  \hat{\sigma}_2 &=& \sin{\theta} d\chi +\cos{\theta} \cos{\chi}
    \sigma_1 +\cos{\theta} \sin{\chi} \sigma_2 -\sin{\theta} \sigma_3,
    \nonumber\\ 
  \hat{\sigma}_3 &=& d\phi +\cos{\theta} d\chi -\sin{\theta} \cos{\chi}
    \sigma_1 -\sin{\theta} \sin{\chi} \sigma_2 -\cos{\theta} \sigma_3.
    \nonumber
\eea
The $\hat{\sigma}_i$ thus contain invariant one-forms for both the SU(2)
group action and the induced rotational action.

\np This is a satisfying result. The metric splits into two parts. One
part describes the interaction of the \two-monopole. It interacts as
if it was a 2-monopole. The other part describes the interaction of
the \one-monopole with the \two-monopole. It is just a Taub-NUT
metric and describes the point dyon interaction of two distinct
monopoles. It should be emphasised that there are order $r^{-2}$
corrections to this approximate metric.                                        

\subsection{The point dyon metric}
\label{ptdymet}

\ns It was pointed out by Atiyah and Hitchin \cite{AH} that for large
separation of the two monopoles, their metric is approximated with
exponential accuracy by a singular Taub-NUT metric. It was
subsequently demonstrated \cite{Ma} that this asymptotic metric could
also have been derived by examining the interactions of point sources
of the fields. This point dyon method was generalised to SU(2)
multimonopoles \cite{GM} and in \cite{LWY3} to larger groups. Although
\cite{LWY3} is mostly concerned with (1,1,\ldots,1)-monopoles, it
describes what the point dyon metric is for any monopole. The point
dyon metrics are discussed in a rigorous way by Bielawski
\cite{Bi1,Bi2}.

\np The radial part of this dyonic metric is 
\begin{equation}
\left(m+\frac{1}{2r_1}+\frac{1}{2r_2}\right)dr^2+\left(\frac{1}{r_2}-\frac{1}{r_1}\right)drdR+\left(2M-\frac{2}{R}+\frac{1}{2r_1}+\frac{1}{2r_2}\right)dR^2
\label{ptdyrmet}
\end{equation}
where $R$ is the distance of each of the \onel-monopoles from the
origin, $r$ is the distance of the \one-monopole from the origin and
$r_1$ and $r_2$ are the distances of each of the \onel-monopoles
from the  \one-monopole. $R$ is large. In this Section, we derive
this metric from the $d\alpha_1$ and  $d\alpha_2$
terms of the $(2,1)$ metric to exponential accuracy in $R$. To do this
we choose a specific configuration and only allow $r$ and $R$ to vary.
This configuration is illustrated in Figure 3. In this
configuration 
\begin{eqnarray}
r_1&=&r-R,\\
r_2&=&r+R.\nonumber
\end{eqnarray}
We have chosen a region where $r$ is bigger than $R$ and the three
monopoles are collinear. This is a convenient choice; however, the
point dyon metric does not rely on $r$ being large \cite{Bi2}.

\begin{figure}[t]
\begin{center}
\begin{picture}(170,50)(35,600)
\thinlines
\put( 45,620){\line( 1, 0){130}}
\put( 35,620){\line( -1, 0){20}}
\put(178,617){\makebox(0,0)[lb]{\smash{\SetFigFont{12}{14.4}{rm}$x_3$-axis}}}
\put( 110,600){\line( 0,+1){ 60}}
\put( 113,657){\makebox(0,0)[lb]{\smash{\SetFigFont{12}{14.4}{rm}$x_1x_2$-plane}}}
\put(  70,620){\circle*{10}}
\put( 150,620){\circle*{10}}
\put( 40,620){\circle{10}}
\put( 70,628){\vector(1,0){  39}}  
\put( 70,628){\vector(-1,0){  0}} 
\put( 87,630){\makebox(0,0)[lb]{\smash{\SetFigFont{12}{14.4}{rm}$R$}}}
\put( 111,628){\vector(1,0){  39}}  
\put( 111,628){\vector(-1,0){  0}} 
\put( 127,630){\makebox(0,0)[lb]{\smash{\SetFigFont{12}{14.4}{rm}$R$}}}
\put( 40,612){\vector(1,0){  69}}  
\put( 40,612){\vector(-1,0){  0}} 
\put( 77,603){\makebox(0,0)[lb]{\smash{\SetFigFont{12}{14.4}{rm}$r$}}}
\end{picture}
\caption{The monopole configuration for the point dyon metric
  calculation, the solid dots represent \onel-monopoles, the other dot
  is the \one-monopole.}
\end{center}
\label{dymetfig}
\end{figure}
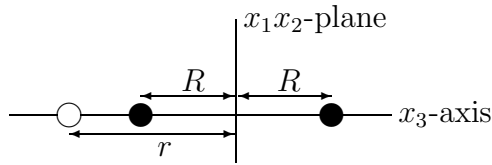

\np The separations $r$ and $R$ must be related to quantities appearing in
the metric. This is easy for $r$, since it is well defined in the full metric,
it is
\begin{equation}
r=-\frac{f_3}{2}.
\end{equation}
Common sense, comparison with the Atiyah-Hitchin case and the numerical calculations of Dancer and Leese
\cite{DL2} all suggest that, up to exponential corrections,
\begin{equation}
R\approx\frac{D}{2}.
\end{equation}
That this suggestion is correct is demonstrated by the success of the
calculation. 
If $R$ is large, $D$ is large. Since $DM<2K$, this implies $2K$ is large. For
large $K$,
\begin{equation}
K\approx\log{4/k'}
\end{equation}
and so $k'$ is exponentially small \cite[eq.600.05]{BF}. The rest of this
Section is concerned with approximating the metric to leading order
in $k'$.

\np In order to evaluate the two integrals $g_1$ and $g_2$ we must
evaluate the incomplete elliptic integral of the second kind,
$E(u;k)$, for $u=DM$. It follows from the quasiperiodicity of
$E(u;k)$, \cite[eqs.113.02 \& 903.01]{BF}, that to leading order in $k'$
\begin{eqnarray}
E(DM;k)&\approx& 2-\sn{k}{DM}\\
       &=&2+\frac{D}{f_3} \nonumber\\
       &\approx&2-\frac{R}{r}.\nonumber
\end{eqnarray}
Now $g_1$ can be evaluated. In Glaisher's notation, see for example
\cite[eq.120.02]{BF},
\begin{equation}
g_1=\frac{1}{D^3}\int_0^{DM}\gn{k}{u}{sd}\,du.
\end{equation}
Now \cite[eq.318.02]{BF} 
\begin{equation}
\int \gn{k}{u}{sd}\,du =
\frac{1}{k^2k'^2}{}\left[E(u;k)-k'^2u-k^2\sn{k}{u}\gn{k}{u}{cd}\right]
\end{equation}
and the Jacobi functions at $u=DM$ can be re-expressed in terms
of $f_1$, $f_2$ and $f_3$. These in turn can be written in terms of
$r$ and $R$ and it can seen that
\begin{equation}
g_1\approx \frac{1}{4R^3 k'^2}.
\end{equation}
$g_2$ can be calculated in a similar way and in fact
\begin{equation}
g_2 \approx \frac{1}{8R^3}\left(2RM-2+\frac{R}{r}\right).
\end{equation}

\np We need to write $d\alpha_1$ and $d\alpha_2$ in terms of
$dr$ and $dR$. Since
\begin{eqnarray}
\alpha_1&=&-k'^2D^2,\\
\alpha_2&=&-D^2,\nonumber
\end{eqnarray}
this can be done if $dk'$ is calculated. To do this we consider
\begin{eqnarray}
-2dr&=&df_3\\
    &=&d(-D\gn{k}{DM}{ns})\nonumber\\
    &=&-D\frac{\partial\gn{k}{DM}{ns}}{\partial
    k}dk-D\frac{\partial\gn{k}{DM}{ns}}{\partial D}dD-\gn{k}{DM}{ns}dD\nonumber
\end{eqnarray}
and use \cite[eq.710.56]{BF}
\be
\frac{\partial \gn{k}{u}{ns}}{\partial k} =
\frac{\gn{k}{u}{cs}\gn{k}{u}{ds}}{kk'^2}\left[E(u;k)-k'^2u-k^2\sn{k}{u}\gn{k}{u}{cd}\right].
\ee
This gives
\begin{eqnarray}
d\alpha_1&\approx&-\frac{1}{2g_1(r^2-R^2)}\left(2dr-\frac{r^2-R^2}{R}MdR-2\frac{r}{R}dR\right)-8k'^2RdR\\
d\alpha_2&\approx&-8RdR\nonumber
\end{eqnarray}
and so, while $g_1$ is order $1/k'^2$, $d\alpha_1$ is order
$k'^2$. The asymptotic metric (\ref{ptdyrmet}) can now be recovered by
substitution.

\section{Discussion}
\label{discusssect}
\news

\begin{figure}[t]
\begin{center}
\begin{picture}(205,120)(85,580)
\thinlines

\put(100,600){\line( 1, 0){ 60}}
\put(100,660){\line( 1, 0){ 60}}
\put(160,640){\line( 1, 0){ 70}}

\put(101,680){\vector( 1,0){  58}}  
\put(101,680){\vector(-1,0){  0}} 
\put(128,683){\makebox(0,0)[lb]{\smash{\SetFigFont{12}{14.4}{rm}$M$}}}

\put(161,680){\vector( 1,0){  68}}  
\put(161,680){\vector(-1,0){  0}}
\put(193,683){\makebox(0,0)[lb]{\smash{\SetFigFont{12}{14.4}{rm}$m$}}}

\put(230,580){\line( 0,1){  100}}
\put(100,580){\line( 0,1){  100}}
\put(160,580){\line( 0,1){  100}}

\put( 97,572){\makebox(0,0)[lb]{\smash{\SetFigFont{12}{14.4}{rm}$s_1$}}}
\put(157,572){\makebox(0,0)[lb]{\smash{\SetFigFont{12}{14.4}{rm}$s_2$}}}
\put(227,572){\makebox(0,0)[lb]{\smash{\SetFigFont{12}{14.4}{rm}$s_3$}}}

\end{picture}
\end{center}
\caption{$(2,1)$ Hanany-Witten configuration with vertical fivebranes
  and horizontal threebranes.}
\label{HWfig}
\end{figure}
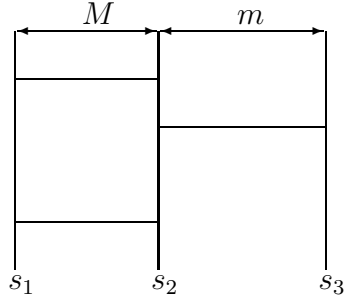

\news \ns We began this calculation with two main motivations. Firstly,
an explicit expression for the $(2,1)$ metric will be useful in the
context of Hanany-Witten theory \cite{HW}; a $(2,1)$-monopole
corresponds to the Hanany-Witten brane configuration illustrated in
Figure 4. Secondly,
it would also be interesting if the $(2,1)$ metric or the $(1,2,1)$
metric could be used to conjecture an approximate form for the metric
of an SU(2) 3-monopole, when two of the monopoles are close together
and the other is far away.

\np If in a SU(2) 3-monopole, two of the monopoles are close together
and one is far away, then the distant monopole has a well defined
position and the two monopoles which are close together have a well
defined centre of mass. This situation is not unlike that considered
above. Figure 5 illustrates this likeness within the Hanany-Witten
notation. In this notation, monopoles are replaced by threebranes which
end on fivebranes. There are $N$ fivebranes for an SU($N$) monopole.
The three dimensions of space are the codimensions in the fivebrane of
the ends of the threebrane.

\begin{figure}[t]

\begin{center}

\begin{picture}(410,140)(120,560)
\thinlines


\put(120,590){\line( 1, 0){ 60}}
\put(120,610){\line( 1, 0){ 60}}
\put(180,660){\line( 1, 0){ 60}}

\put(121,680){\vector( 1,0){  58}}  
\put(121,680){\vector(-1,0){  0}} 
\put(143,684){\makebox(0,0)[lb]{\smash{\SetFigFont{12}{14.4}{rm}$M$}}}

\put(181,680){\vector( 1,0){  58}}  
\put(181,680){\vector(-1,0){  0}}
\put(203,684){\makebox(0,0)[lb]{\smash{\SetFigFont{12}{14.4}{rm}$M$}}}

\put(240,580){\line( 0,1){  100}}
\put(120,580){\line( 0,1){  100}}
\put(180,580){\line( 0,1){  100}}


\put(300,590){\line( 1, 0){ 60}}
\put(300,610){\line( 1, 0){ 60}}
\put(360,660){\line( -1, 0){ 60}}

\put(301,680){\vector( 1,0){  58}}  
\put(301,680){\vector(-1,0){  0}} 
\put(323,684){\makebox(0,0)[lb]{\smash{\SetFigFont{12}{14.4}{rm}$M$}}}

\put(300,580){\line( 0,1){  100}}
\put(360,580){\line( 0,1){  100}}


\put(450,590){\line( 1, 0){ 60}}
\put(450,610){\line( 1, 0){ 60}}
\put(510,660){\line( 1, 0){ 30}}
\put(450,660){\line( -1, 0){ 30}}

\put(451,680){\vector( 1,0){  58}}  
\put(451,680){\vector(-1,0){  0}} 
\put(473,684){\makebox(0,0)[lb]{\smash{\SetFigFont{12}{14.4}{rm}$M$}}}

\put(511,680){\vector( 1,0){  28}}  
\put(511,680){\vector(-1,0){  0}}
\put(513,684){\makebox(0,0)[lb]{\smash{\SetFigFont{12}{14.4}{rm}$M/2$}}}

\put(421,680){\vector( 1,0){  28}}  
\put(421,680){\vector(-1,0){  0}}
\put(423,684){\makebox(0,0)[lb]{\smash{\SetFigFont{12}{14.4}{rm}$M/2$}}}

\put(420,580){\line( 0,1){  100}}
\put(540,580){\line( 0,1){  100}}
\put(450,580){\line( 0,1){  100}}
\put(510,580){\line( 0,1){  100}}

\put(172,560){\makebox(0,0)[lb]{\smash{\SetFigFont{12}{14.4}{rm}(a)}}}
\put(322,560){\makebox(0,0)[lb]{\smash{\SetFigFont{12}{14.4}{rm}(b)}}}
\put(472,560){\makebox(0,0)[lb]{\smash{\SetFigFont{12}{14.4}{rm}(c)}}}

\end{picture}

\end{center}

\caption{Hanany-Witten configurations illustrating the discussion of
  two monopoles close together and one far away. (a) corresponds to a
  $(2,1)$-monopole where $m=M$. (b) corresponds to a 3-monopole, the
  topmost threebrane is far away from the other two. (c) corresponds
  to a $(1,2,1)$-monopole.}

\end{figure}
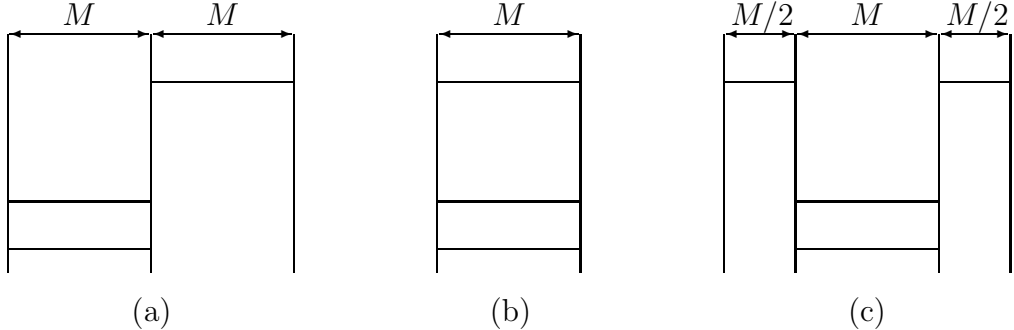

\np There are two types of interaction between the threebranes. There is a
short range force acting between the whole length of vertically
aligned threebranes and a long range force acting on the ends of the
threebranes through the fivebranes. In Figure 5(a) the \two-threebranes
interact through both types of force, the \one-threebrane interacts
with the \two-threebranes only through the fivebrane upon which they
both end. In Figure 5(b), there are two threebranes close together and
one far away. To an exponential approximation, the far away threebrane
only interacts with the other threebranes through the two fivebranes
upon which they all end.  Because of this, it might be expected that
the moduli space of 3-monopoles with two monopoles close together and
one far away, is exponentially well approximated by the $(2,1)$ metric.

\np In the 3-monopole, the threebranes end on the fivebranes from the
same side and they all end on two different fivebranes. In the
$(2,1)$-monopole, the \two-threebranes end on one side of a fivebrane
and the \one-threebrane on the other. This implies that for the
approximation to work a sign and a factor of two must be
changed in the $(2,1)$ metric. In the point dyon metric calculation of
Section \ref{ptdymet}, the correct change of sign results if $DM$ is
greater than $2K$, rather than less than it. The factor of two would be
corrected if, instead of the $(2,1)$ metric, the $(1,2,1)$ metric was
used with the $(1,\;,\;)$-monopole constrained to be coincident with the
$(\;,\;,1)$-monopole as illustrated in Figure 5(c). The construction
of the $(1,2,1)$ metric is discussed below. This picture is very vague,
a more precise understanding of the relationship between asymptotic SU(2)
metrics and SU($N$) metrics would require the more sophisticated
methods found in  \cite{Bi1,Bi2}.

\np We were also motivated to calculate the $(2,1)$ metric by \cite{Ch}.
Among the calculations in this paper, is an attempt to calculate the
$(2,1)$ metric using the Legendre transform construction \cite{HKLR,IR}.
The constraint equations arising in the Legendre transform prove
intractable but the formulation itself is of great interest. Even
without solving the constraint equation, Chalmers is able to study
some features of the $(2,1)$ metric, for example, he is able to extract
the point dyon metric.  We hope that our work will prove useful in
investigating the Legendre transform construction of the $(2,1)$ metric.

\np An interesting aspect of our calculation is the compelling form of
the relative phase coordinate (\ref{relphase}). It is the gauge
invariant combination of the phase coordinates for the \two-monopole
and the \one-monopole. In Section \ref{HKQsect}, this phase
coordinate arises naturally when we use the hyperK\"ahler quotient
construction to construct the $(2,1)$ metric by attaching a 1-monopole
to a $(2,[1])$-monopole.  This use of the hyperK\"ahler construction
provides a simple method for constructing monopoles with no more than
two monopoles of any type. In order to complete the tool box for such
constructions, we have calculated the $([1],2,[1])$ metric in Section
\ref{121sect}. From this, the $(1,2,1)$ metric could be
calculated by choosing two suitable U(1) actions. As another
example, three U(1) actions could be used to construct the $(2,1,2,1)$
metric, by attaching a 1-monopole between the $(2,[1])$ metric and the
$([1],2,1)$ metric. These methods could also be used to calculated
moduli space of monopoles with other gauge groups like SO(5).  Calculating the
$(2,2)$ metric by attaching the $(2,[1])$ metric to the $([1],2)$ metric
should not be too difficult either. It would simply require a larger
action with which to perform the hyperK\"ahler quotient.

\np The analysis of Section \ref{submansect} suggests that it is unlikely
that there are bound geodesics on $\MM{0}$.  An investigation of the
quantum mechanics on $\MM{0}$ would be interesting but very difficult.
The classical behaviour suggests the absence of bound states. In
addition, S-duality predicts the absence of a Sen form on this space,
because the form would be dual to W-bosons that are not seen at low
energies \cite{GL,LWY3}.

\section{A space of SU(4) monopoles}
\label{121sect}
\news

\ns In this Section, the metric on a space of $([1],2,[1])$-monopoles is
calculated. The boundary condition breaks the  symmetry from SU(4) to
SU(2)$\times$U(1)$\times$SU(2). The
asymptotic Higgs field lies in the gauge orbit of
\begin{equation}
\Phi_{\infty}=\left(\begin{array}{cccc}s_1&&&\\&s_1&&\\&&s_2&\\&&&s_2
\end{array}\right).
\end{equation}
where $s_2=-s_1$ and for convenience we choose $s_1=-2$.  Roughly
speaking, a $([1],2,[1])$-monopole is composed of two massive monopoles
and two massless monopoles. The two massive monopoles are of the same
type and are of different type to the two massless monopoles. These
massless monopoles are of different types to each other.  The
relative moduli space $\MG{[1],2,[1]}$ is twelve dimensional, these
twelve correspond to the separation of the massive monopoles, their
SO(3) orientation in space, two sets of three SU(2) parameters
corresponding to the unbroken gauge group and finally, two cloud
parameters.  In a $(1,2,1)$-monopole the $(1,\;,\;)$-monopole does not
interact with the $(\;,\;,1)$-monopole except in so far as each of
them affects the $(\;,2,\;)$-monopole.  Since the two clouds in the
$([1],2,[1])$ case correspond to a massless $(1,\;,\;)$-monopole and a
massless $(\;,\;,1)$-monopole they should interact in a relatively
simple manner.

\np In the limit, when one of
the clouds is at infinity, $\MG{[1],2,[1]}$ reduces to $\MD$ with an
additional infinite term. If both clouds are at infinity, it reduces to
Atiyah-Hitchin but with two infinite terms. A restriction which amounts
to identifying the two clouds, reduces the metric to the one
considered in \cite{LL}. 

\np The general features of this SU(4) metric were discussed in
\cite{H}. The family of \hK four-manifolds
which are the \hK quotients of $\MG{[1],2,[1]}$ were discussed
in \cite{D3,H}. Along with the previous calculation, the ease with which
the monopole metric is found, once the Nahm equations are solved, shows
that the monopole metric problem for larger groups is no more
difficult than the problem when the group is SU(2).

\np The $([1],2,[1])$ Nahm data are $2\times2$ skewHermitian matrix
functions of $s\in[-2,2]$. They satisfy the Nahm equation and are
analytic over the entire closed region. They are represented by the
diagram
\begin{equation}
\begin{array}{c}
\begin{picture}(205,65)(55,615)
\thinlines
\put( 60,620){\line( 1, 0){200}}
\put(100,620){\line( 0, 1){ 40}}
\put(100,660){\line( 1, 0){ 60}}
\put(100,640){\line( -1, 0){ 2}}
\put( 98,620){\line( 0, 1){ 20}}
\put(160,660){\line( 0,-1){ 40}}
\put(160,640){\line( 1,0){  2}}
\put(162,640){\line( 0,-1){ 20}}
\put( 93,623){\vector(0,1){  36}}  
\put( 93,623){\vector(0,-1){  0}} 
\put(100,620){\line( 0,-1){  5}}
\put(160,620){\line( 0,-1){  5}}
\put( 82,638){\makebox(0,0)[lb]{\smash{\SetFigFont{12}{14.4}{rm}2}}}
\put( 91,605){\makebox(0,0)[lb]{\smash{\SetFigFont{12}{14.4}{rm}$-2$}}}
\put(157,605){\makebox(0,0)[lb]{\smash{\SetFigFont{12}{14.4}{rm}$2$}}}
\end{picture}\end{array}
\end{equation}
The gauge action is $G^{0,0}$. There are two SU(2) actions, one given
by $G^{0,\star}/G^{0,0}$ and the other by $G^{\star,0}/G^{0,0}$. There
is also a translational action, like the one discussed above and a
rotational action given by
\begin{eqnarray}
T_0&\rightarrow& T_0,\\
T_i&\rightarrow& \,\sum_jR_{ij}T_j.
\end{eqnarray}
The centre of mass and the U(1) action given by
$\exp({if(s){\bf1}_2})$ are  fixed in the same way as before. This means that
the Nahm data are traceless.

\np As before, the
Nahm equations are reduced by the ansatz
\begin{eqnarray}
&&T_0(s)=0,\\
&&T_i(s)=f_i(s)e_i\nonumber
\end{eqnarray}
to the Euler-Poinsot top equations. The solutions are Euler top
functions but with weaker boundary conditions:
\begin{eqnarray}
f_1(s)&=&\pm\frac{Dcn_kD(s+\tau)}{sn_kD(s+\tau)},\\
f_2(s)&=&\pm\frac{Ddn_kD(s+\tau)}{sn_kD(s+\tau)},\nonumber\\
f_3(s)&=&\pm\frac{D}{sn_kD(s+\tau)},\nonumber
\label{1.4}
\end{eqnarray}
with 0$\leq k\leq$ 1, $\,\tau>$2 and $D(\tau+2)<\,$2$K(k)$. The extra
parameter $\tau$ reflects the analyticity of the Nahm data at $s=-2$.
All signs are negative or exactly two signs are positive. More solutions
are given by sending $s$ to $-s$ and changing the signs of $f_i$. 

\np Using these solutions to the Nahm equations, it is a
simple exercise to find the metric
on ${\cal M}_{([1],2,[1])}$. The metric depends on $D,k$ and $\tau$ 
and three sets of SU(2) coordinates. Each of the SU(2)'s
acts isometrically, so it is only necessary 
to calculate the metric in the neighbourhood of the
identity of the SU(2)'s. The SU(2) actions are then used to
find the metric at a general point. The method is a duplicate of that used 
in \cite{D,I} and  the interested reader is referred to these papers.

Coordinates on the quotient space
$\MG{[1],2,[1]}/($SO(3)$\times$SU(2)$\times$SU(2)$)$ are provided by
\begin{eqnarray}
\alpha_1&=&k'^2 D^2,\\
\alpha_2&=&D^2,\nonumber\\
\alpha_6&=&\sum_i[f_i^2(2)-f_i^2(-2)].\nonumber
\end{eqnarray}
The remaining nine
coordinates on $\MG{[1],2,[1]}$ are given by SU(2) matrices
corresponding to rotations $R\inn$~SO(3) and group actions given by
$G(2)\inn$~SU(2) for $G^{0,\star}/G^{0,0}$ and $G(-2)\inn$~SU(2) for
$G^{\star,0}/G^{0,0}$. The metric is written in terms of
left-invariant one-forms given by
\begin{eqnarray}
\frac{i}{2}\tau_i\sigma_i&=&R^{\dagger}dR,\\
\frac{i}{2}\tau_i\check{\sigma}_i&=&G(2)^{\dagger}dG(2),\nonumber\\
\frac{i}{2}\tau_i\hat{\sigma}_i&=&G(-2)^{\dagger}dG(-2).\nonumber
\end{eqnarray}
where it should be noted that $\check{\sigma}$ and $\hat{\sigma}$ are
different symbols.
Because the SU(2) actions are isometric, the coefficients of the metric
depend only on $D$, $k$ and $\tau$.  In order to express the metric, it
is useful to define the following functions of $D$, $k$ and $\tau$;
\begin{eqnarray}
g_1(k,D,\tau)&=&\int_{-2}^2\frac{1}{f_2^2}\,ds\,,\quad\qquad\;\;
g_2(k,D,\tau)=\int_{-2}^2\frac{1}{f_3^2}\,ds\,,\\
\nonumber\\
A(k,D,\tau)&=&f_1(2)f_2(2)f_3(2)\,,\quad
B(k,D,\tau)=f_1(-2)f_2(-2)f_3(-2)\,,\nonumber\\
\nonumber\\
X(k,D,\tau)&=&AB(g_1+g_2)+B-A\,.\nonumber
\end{eqnarray}
The following combination of one-forms also simplifies the expression for
the metric
\begin{eqnarray}
\omega_1&=&\check{\sigma}_1-\frac{f_3(2)}{f_2(2)}\sigma_1,\\
\omega_2&=&\check{\sigma}_2-\frac{f_1(2)}{f_3(2)}\sigma_2,\nonumber\\
\omega_3&=&\check{\sigma}_3-\frac{f_1(2)}{f_2(2)}\sigma_3\nonumber
\end{eqnarray}
and
\begin{eqnarray}
\rho_1&=&\hat{\sigma}_1-\frac{f_3(-2)}{f_2(-2)}\sigma_1,\\
\rho_2&=&\hat{\sigma}_2-\frac{f_1(-2)}{f_3(-2)}\sigma_2,\nonumber\\
\rho_3&=&\hat{\sigma}_3-\frac{f_1(-2)}{f_2(-2)}\sigma_3.\nonumber
\end{eqnarray}
The metric is then given by the following complicated expression:
\begin{eqnarray}\label{1.8}
ds^2&=&\frac{AB}{4(B-A)}(g_1d\alpha_1+g_2d\alpha_2)^2
+\frac{1}{4}(g_1d\alpha_1^2+g_2d\alpha_2^2)
+\frac{1}{A-B}d\alpha_6^2 \\
&+&\frac{g_1(ABg_2+B-A)k^4D^4}{X}\sigma_1^2
+\frac{(g_1+g_2)g_2D^4}{g_1}\sigma_2^2
+\frac{(g_1+g_2)g_1k'^4D^4}{g_2}\sigma_3^2 \nonumber\\
&+&\frac{A^2(B(g_1+g_2)-1)}{f_1(2)^2X}\omega_1^2
+\frac{Ag_1+1}{f_2(2)^2g_1}\omega_2^2+\frac{Ag_2+1}{f_3(2)^2g_2}
\omega_3^2\nonumber\\
&-&\frac{B^2(A(g_1+g_2)+1)}{f_1(-2)^2X}\rho_1^2
-\frac{Bg_1-1}{f_2(-2)^2g_1}\rho_2^2
-\frac{Bg_2-1}{f_3(-2)^2g_2}\rho_3^2\nonumber\\
&+&\frac{2ABg_1k^2D^2}{f_1(2)X}\sigma_1\omega_1
+\frac{2D^2g_2}{g_1f_2(2)}\sigma_2\omega_2
+\frac{2D^2g_1k'^2}{g_2f_3(2)} \sigma_3\omega_3\nonumber\\
&-&\frac{2ABg_1k^2D^2}{f_1(-2)X}\sigma_1\rho_1
-\frac{2D^2g_2}{g_1f_2(-2)}\sigma_2\rho_2
-\frac{2D^2g_1k'^2}{g_2f_3(-2)}\sigma_3\rho_3\nonumber\\
&+&\frac{2AB}{f_1(2)f_1(-2)X}\omega_1\rho_1
-\frac{2}{g_1f_2(2)f_2(-2)}\omega_2\rho_2
-\frac{2}{g_2f_3(2)f_3(-2)}\omega_3\rho_3.\nonumber
\end{eqnarray}

\np The Sp(4) condition \cite{LL} can be imposed: $f_1(2)=-f_1(-2)$,
$f_2(2)=f_2(-2)$ and $f_3(2)=f_3(-2)$, along with the identifications
$\check{\sigma}_1=\hat{\sigma}_1$, $\check{\sigma}_2=-\hat{\sigma}_2$,
$\check{\sigma}_3=-\hat{\sigma}_3$ . This reduces (\ref{1.8}) to the metric
found in \cite{LL}.  Alternatively, the limit $A\rightarrow\infty$ or
$B\rightarrow\infty$ can be taken and it can be shown that this reduces
the $([1],2,[1])$ metric to the $(2,[1])$ metric discussed in
\cite{D,I}, along with an infinite term corresponding to the moment of
inertia of the cloud at infinity. If $A\rightarrow\infty$ and
$B\rightarrow\infty$, the metric reduces to the Atiyah-Hitchin metric
along with infinite terms corresponding to the inertia of both clouds at
infinity. 

\np There is a three-dimensional geodesic submanifold of ${\cal
  M}^{([1],2,[1])}$ obtained by imposing $D_2$ symmetry on the
monopoles. This was denoted $\X$ in \cite{H} and is described
there. It is the $([1],2,[1])$ analogue of the $Y$ space of Dancer and
Leese \cite{DL1}. Imposing spherical symmetry on $\X$ reduces to
one-dimensional submanifolds whose Nahm data are
\begin{equation}\label{sphers} 
f_1(s)=f_2(s)=f_3(s)=-\frac{1}{s+\tau} 
\end{equation} 
where $\tau>2$ or $\tau<-2$. Using the expression for the metric, we
can determine the geodesics in this one-dimensional example.  If $\tau>2$
and initially decreasing, then it continues to approach $\tau$=2
without ever reaching there. This corresponds to one of the clouds
increasing to arbitrarily large radius with the other cloud remains
small. The massive monopoles are at the origin. If $\tau>2$ and
initially increasing, then it reaches $\tau=\infty$ in finite time and
then, since $\tau=\infty$ and $\tau=-\infty$ are equivalent by
(\ref{sphers}), $\tau$ increases from $-\infty$ to approach $-2$. If
$\tau$ is initially close to two, this represents a process where one
cloud is initially large and decreasing. It reaches its minimum size
and then the other cloud continually increases from its minimum size
to arbitrarily large radius.

\section*{Acknowledgements}
CJH benefited from discussions with Paul M. Sutcliffe and is grateful to
Fitzwilliam College, Cambridge for support. PWI thanks NSERC of Canada
and FCAR of Qu\'ebec for financial assistance. AJM thanks PPARC and the
1851 Royal Commission for financial support.

\end{document}